\documentclass{aa}
\usepackage{graphicx}
\usepackage{placeins}
\usepackage[varg]{txfonts}
\usepackage{multirow}
\usepackage{float}
\usepackage{array}
\usepackage{lscape}
\bibpunct{(}{)}{;}{a}{}{,}
\usepackage{natbib,twoopt}
\usepackage[breaklinks=true]{hyperref}
\usepackage{orcidlink}
\definecolor{linkblue}{rgb}{0,0.4,0.6}
\hypersetup{colorlinks, linkcolor={linkblue}, citecolor={linkblue}, urlcolor={linkblue}}

\title{Cosmic evolution of the [C\,\textsc{ii}]-to-dense neutral gas relation}
\titlerunning{Cosmic evolution of the [C\,\textsc{ii}]-to-dense neutral gas relation}

\author{
    C. Accard           \corrauth{cedric.accard@astro.unistra.fr}      \inst{\ref{ObAS}}   \orcidlink{0009-0005-9982-7239} 
    \and F. Renaud      \email{florent.renaud@astro.unistra.fr}     \inst{\ref{ObAS}}   \orcidlink{0000-0001-5073-2267}
    \and K. Kraljic     \email{katarina.kraljic@astro.unistra.fr}   \inst{\ref{ObAS}}   \orcidlink{0000-0001-6180-0245}
    \and D. Ismail      \email{diana.ismail@astro.unistra.fr}       \inst{\ref{ObAS}}   \orcidlink{0009-0007-2281-4944}
    \and M. Béthermin   \email{matthieu.bethermin@astro.unistra.fr} \inst{\ref{ObAS}}   \orcidlink{0000-0002-3915-2015}
    \and O. Agertz      \email{oscar.agertz@fysik.lu.se}           \inst{\ref{Lund}}   \orcidlink{0000-0002-4287-1088}
    }

\institute{
    Université de Strasbourg, CNRS, Observatoire Astronomique de Strasbourg, UMR 7550, 67000 Strasbourg, France \label{ObAS}
    \and 
    Lund Observatory, Division of Astrophysics, Department of Physics, Lund University, Box 43, SE-22100 Lund, Sweden \label{Lund}}
    
\date{Received 21 April 2026 / Accepted 24 August 2026}

\begin{document}

\abstract
{The [C\,\textsc{ii}] 158\,$\mu$m line is widely used to trace star formation and the gas contents of high-redshift galaxies. However, it remains unclear under which physical conditions it reliably traces the molecular reservoir, and whether a unique conversion factor $\alpha_{\rm [C\,\textsc{ii}]}$ can be applied across cosmic time. We investigated the evolution of the relation between the [C\,\textsc{ii}] luminosity and dense neutral gas mass from $z\simeq10$ to $z\simeq0.2$ using the \textsc{Vintergatan} simulation, a high-resolution cosmological zoom-in of a Milky Way–like galaxy. We post-processed the snapshots with the \textsc{Skirt} radiative transfer code to generate synthetic [C\,\textsc{ii}] data cubes. We measured global and spatially resolved (100\,pc) relations between [C\,\textsc{ii}] luminosity ($L_{\rm [C\,\textsc{ii}]}$), star formation rate (SFR), and dense neutral gas mass ($M_{\rm DN}$). We followed the redshift evolution of the [C\,\textsc{ii}]–to-dense neutral gas conversion factor $\alpha_{\rm [C\,\textsc{ii}]}$, and linked these trends to the evolution of the interstellar medium (ISM) phases. The global $L_{\rm [C\,\textsc{ii}]}$–$M_{\rm DN}$ and $L_{\rm [C\,\textsc{ii}]}$–SFR relations evolve from a steep, [C\,\textsc{ii}]-deficient regime at a very low metallicity to an almost linear behaviour, similar to calibrations at $z\approx2$, once the ISM reaches $Z_{\rm gas}\gtrsim0.05$–$0.1\,Z_\odot$ at $z\lesssim5$. Over this evolution, $\alpha_{\rm [C\,\textsc{ii}]}$ spans nearly three orders of magnitude, from $\gtrsim10^4$ down to $\approx10\,\rm{M_\odot\,L_\odot^{-1}}$, even though the [C\,\textsc{ii}] emission remains spatially correlated with the dense neutral gas. We provide prescriptions for $\alpha_{\rm [C\,\textsc{ii}]}$ based on integrated and local quantities. The latter recovers the dense neutral mass of \textsc{Vintergatan} within factors of the of order unity, whereas global relations depending only on $L_{\rm [C\,\textsc{ii}]}$ and/or the integrated metallicity fail by up to an order of magnitude during major and minor merger epochs. A unique, redshift-independent $\alpha_{\rm [C\,\textsc{ii}]}$ therefore cannot recover dense neutral gas masses across the regimes we explored. [C\,\textsc{ii}] remains a viable tracer of dense neutral gas at very high redshifts, but only when used with conversion factors that explicitly account for metallicity, ISM phase mix, and merger events.}


\maketitle
\nolinenumbers

\section{Introduction}\label{sec:intro}

A central goal of galaxy evolution studies is to understand how baryons cycle between gas and stars across cosmic time \citep[e.g.][]{Madau2014, Walter2020}. The cosmic star formation rate (SFR) density rises rapidly from the earliest epochs, peaks around $z\approx 2$, and subsequently declines, reflecting a complex interplay between gas accretion, star formation, and feedback \citep[e.g.][]{Bouche2010}. In this context, the small-scale physics of the interstellar medium (ISM), its density structure, phase mix, together with dynamics regulate how efficiently galaxies can convert their gas reservoirs into stars at different epochs and in different environments \citep[e.g.][]{Krumholz2012,Hopkins2012,Krumholz2018}. Limited sensitivity, coarse spatial resolution, and heterogeneous samples still prevent us from establishing whether early galaxies simply follow the same relations for star formation as in the local Universe, or instead host a qualitatively different star-forming regime \citep[e.g.][]{Genzel2010,Tacconi2013,Freundlich2013}. In this context it is essential to understand how to infer molecular gas masses from observables \citep[e.g.][]{Tacconi2020}.

At a low redshift, the most widely used molecular gas tracer is low‑$J$ CO, calibrated via a CO‑to‑$\rm H_2$ conversion factor $\alpha_{\rm CO}$ that depends on the environment, in particular the metallicity and dynamical state \citep[e.g.][]{Bolatto2013}. In nearby galaxies, $\alpha_{\rm CO}$ increases sharply below $Z_{\rm gas} \approx 0.3 - 0.5 \, \rm{Z}_\odot$ as CO becomes confined to the densest, best‑shielded regions, and variations in depletion time correlate more strongly with dynamical conditions than with instantaneous SFR \citep[e.g.][]{Renaud2019}. These results already indicate that, even in the conditions of the local Universe, commonly used gas tracers require environment-dependent conversion factors. At a higher redshift, however, CO observations become progressively more expensive and, beyond $z\gtrsim4$, often impractical for large samples, motivating the use of alternative tracers of the cold gas reservoir such as the [C\,\textsc{ii}] 158$\,\mu$m line \citep[e.g.][]{Rizzo2024,Roman2024}.

The [C\,\textsc{ii}] 158$\,\mu$m fine-structure line is typically among the brightest far-infrared lines in high‑redshift star-forming galaxies and it is redshifted into the Atacama Large Millimeter/submillimeter Array (ALMA) bands for $z\gtrsim4$, making it a cornerstone tracer \citep[e.g.][]{CarilliWalter2013, LeFevre2020, Fudamoto2022}. The line is primarily excited in photodissociation regions (PDRs) and at the surfaces of molecular clouds, and it acts as a major coolant of the cold, neutral ISM. It is thus widely used as a tracer of both star formation and the cold gas reservoir \citep[e.g.][]{Pabst2022, Wolfire2022}. However, [C\,\textsc{ii}] can also be emitted from more diffuse neutral and ionised gas, with the relative contribution depending on the metallicity, radiation field, and gas surface density, which complicates any simple conversion between $L_{\rm [C\,\textsc{ii}]}$ and either the SFR or the molecular gas mass \citep[e.g.][]{Pineda2014, Vallini2015, Accurso2017}. By analogy with CO, one can expect the [C\,\textsc{ii}]–to–gas conversion factor to depend not only on the metallicity but also on the local ISM structure, pressure, and radiation field. Calibrations derived in relatively metal‑rich, star-forming galaxies may not extrapolate straightforwardly to more extreme, low‑metallicity regimes found at high redshifts.

Large ALMA programmes such as ALPINE \citep{LeFevre2020, Bethermin2020} and REBELS \citep{Bouwens2022}, along with targeted follow-up campaigns (e.g., ALPINE-CRISTAL; \citealt{HerreraCamus2025}), have now produced sizeable samples of [C\,\textsc{ii}]-detected galaxies at $z\gtrsim4$. These data enable measurements of gas kinematics, sizes, and dynamical masses, and they allow for the first attempts at deriving resolved [C\,\textsc{ii}]–SFR and [C\,\textsc{ii}]–gas scaling relations at early times \citep[e.g.][]{Schaerer2020, Faisst2020}. While current efforts have established that [C\,\textsc{ii}] can broadly trace star formation and cold gas at a high redshift \citep[e.g.][]{DeLooze2014, Zanella2018}, they also reveal substantial scatter and systematic offsets relative to local [C\,\textsc{ii}]-based SFR and gas calibrations \citep[e.g.][]{Schaerer2020,Accard2025}. In particular, a [C\,\textsc{ii}] deficit (i.e. a drop in $L_{\rm [C\,\textsc{ii}]}/L_{\rm IR}$ or an apparent underluminosity of [C\,\textsc{ii}] at a fixed SFR) appears to emerge in compact starbursts and some high-$z$ main-sequence galaxies, raising questions on how reliably [C\,\textsc{ii}] can be used as a quantitative tracer in the early Universe and whether a single, locally calibrated $\alpha_{\rm [C\,\textsc{ii}]}$ is adequate over a broad range of redshifts and ISM conditions \citep[e.g.][]{Malhotra2001, Lagache2018, Ferrara2019}. A physically grounded revision of these conversion factors must explicitly track how the physical conditions of the [C\,\textsc{ii}]-emitting gas, and the (typically minor) contributions from non-molecular phases, evolve with redshift and with global galaxy properties such as the metallicity and dynamical state \citep[e.g.][]{Narayanan2017}.

Addressing these issues requires information that is difficult or impossible to obtain from observations alone, especially below kiloparsec scales in the early Universe. High-resolution cosmological zoom-in simulations resolve the internal structure of the ISM, the clustered nature of star formation, and the time-variable impact of feedback, providing a self-consistent laboratory for exploring how [C\,\textsc{ii}] and other lines respond to changing physical conditions in galaxies \citep[e.g.][]{Hopkins2014_FIRE,Somerville2015_review}. However, accurately modelling the emergent line emission with full 3D radiative transfer is computationally expensive, particularly at the resolution of state-of-the-art cosmological zoom-in simulations. It is then often applied in post-processing rather than evolved self-consistently on the fly \citep[e.g.][]{Vallini2015,Olsen2021, Schimek2024, Liu2026}. Recent radiation-hydrodynamical efforts, such as the MEGATRON simulations \citep{Katz2025,Choustikov2026} and non-equilibrium thermochemistry schemes for metal ions \citep[e.g.][]{Lupi2020}, take important steps towards predicting line emission self-consistently during the simulation, but they still neglect dust scattering and re-emission. As a result, detailed dust radiative transfer is currently more practical to perform in post-processing than to evolve self-consistently. For this work, we adopted such a post-processing approach, applying 3D radiative transfer to simulation snapshots. This strategy is computationally more tractable than performing full radiation–hydrodynamics at the same resolution, while still capturing the geometry, dust distribution, and radiation fields that shape the emergent [C\,\textsc{ii}] emission \citep[e.g.][]{Baes2011}. This approach enables both a decomposition of [C\,\textsc{ii}] into its contributing ISM phases and the derivation and validation of spatially resolved versus global [C\,\textsc{ii}]-to-gas conversion recipes tailored to high-redshift conditions \citep[e.g.][]{Ferrara2019}.

In this work, we combined the high-resolution cosmological zoom-in simulation \textsc{Vintergatan} \citep{Agertz2021} with the 3D Monte Carlo radiative transfer code \textsc{Skirt} \citep{Camps2020}. We analysed both global and spatially resolved [C\,\textsc{ii}]–SFR and [C\,\textsc{ii}]–gas relations to establish under which conditions [C\,\textsc{ii}] remains a straightforward tracer of the dense neutral phase and, by extension, of the molecular reservoir. To assess the accuracy of global versus resolved $\alpha_{\rm [C\,\textsc{ii}]}$ prescriptions, we provide updated, physically grounded guidelines for interpreting [C\,\textsc{ii}] measurements at high redshift.

The paper is organised as follows. In Section 2 we describe the \textsc{Vintergatan} simulation and our phase classification of the ISM, the radiative-transfer post-processing and the extraction of the [C\,\textsc{ii}]-related quantities. In Section 3 we present the global [C\,\textsc{ii}]–dense neutral gas relation and its redshift evolution, and compare \textsc{Vintergatan} to existing empirical and simulation-based scaling relations. Section 4 discusses the physical origin of the [C\,\textsc{ii}] deficits, the role of different ISM phases, and the implications for using [C\,\textsc{ii}] as a dense neutral gas tracer at high redshift. Our main conclusions and prospects for future work are summarised in Section 5.

\section{Methods} \label{sec:methods}
\subsection{The VINTERGATAN simulation} \label{subsec:vintergatan}

We analysed the \textsc{Vintergatan} project, a high-resolution cosmological zoom-in simulation of a Milky Way–mass galaxy (M$_{\star} \approx 6\times10^{10}\,\mathrm{M_\odot}$ at $z=0.2$). A detailed description of the simulation setup, and sub-grid physics implementation is provided in \citet{Agertz2021}; here we briefly summarise the aspects that are most relevant to the present work.

The simulation was performed with the adaptive mesh refinement (AMR) code \textsc{Ramses} \citep{Teyssier2002}. The adopted cosmology was a flat $\Lambda$CDM model with $H_0= 70.2\,\mathrm{km\,s^{-1}\,Mpc^{-1}},\ \Omega_{\rm m} = 0.272,\ \Omega_{\Lambda} = 0.728$, and $\Omega_{\rm b} = 0.045$. A key feature of \textsc{Vintergatan} is its high numerical resolution, achieved through a pseudo-Lagrangian refinement strategy that follows the densest structures in the ISM. The simulation reaches a spatial resolution of $\Delta x \approx 20$ pc in the highest-density regions. The dark matter mass resolution in the zoom-in region is $3.5\times10^{4}\,\mathrm{M_\odot}$, while the initial gas mass resolution is $7070\,\mathrm{M_\odot}$.

Star formation was modelled in gas above a density threshold $n > 100\,\mathrm{cm^{-3}}$, with a local efficiency per free-fall time that depends on the turbulent properties of the gas \citep{Padoan2012}. Stellar feedback includes the combined effects of stellar winds, radiation pressure, and Type Ia and Type II supernovae. The evolution of the galaxy structure and its chemical enrichment history is discussed in detail in the companion papers \citet{Renaud2021b, Renaud2021a, SegoviaOtero2022, Renaud2025}.

To associate the [C\,\textsc{ii}] emission with specific ISM components, we classified the gas into distinct phases using temperature and density thresholds. Following previous cosmological simulations of diverse galaxy populations, such as \textsc{NewHorizon} \citep{Dubois2021}, we defined the diffuse neutral phase as gas with $T < 2\times10^{4}\,\mathrm{K}$ and $0.1 < n < 10\,\mathrm{cm^{-3}}$, and the dense neutral phase as gas with $T < 2\times10^{4}\,\mathrm{K}$ and $n > 10\,\mathrm{cm^{-3}}$; gas at lower densities ($n \leq 0.1\,\mathrm{cm^{-3}}$) or higher temperatures was grouped into a warm-hot diffuse phase. Because \textsc{Vintergatan} does not explicitly follow the H$_2$ fraction, these thresholds do not provide a direct definition of molecular gas. Instead, the dense neutral phase encompasses gas expected to host the H$_2$ reservoir and is used here as a proxy for it. A self-consistent molecular fraction would require explicit chemistry and radiative transfer \citep[e.g.][]{Lupi2018,Nickerson2019}. We have verified that adopting a more restrictive cold, dense selection ($T<10^{2}\,\mathrm{K}$, $n>10\,\mathrm{cm^{-3}}$) changes the inferred dense neutral-gas mass by less than an order of magnitude and does not alter the redshift evolution or depth of the drop in $\alpha_{\rm [C\,\textsc{ii}]}$ discussed in Sect.~\ref{subsec:resolved_alpha}.

In this work, we compared the synthetic [C\,\textsc{ii}] emission with the SFR of the simulation. The SFR surface density maps ($\Sigma_{\rm SFR}$) were directly computed from the stellar particle population on a grid of 100 pc resolution, chosen for the reasons discussed in Sect.~\ref{subsec:Skirt}. To explore how the correlation between [C\,\textsc{ii}] emission and star formation depends on the averaging timescale, and consequently on different stages of the star formation process, we constructed $\Sigma_{\rm SFR}$ maps over three timescales: $\tau_{\rm SF}=10$, $30$, and $100$ Myr. We adopted $\tau_{\rm SF}=10$ Myr as our fiducial choice, as it corresponds to the timescale on which feedback regulates the local ISM. A comparison with the other timescales, for a subset of the results in Sect.~\ref{sec:results}, is provided in Appendix~\ref{apdx:left_out}.

Throughout this work, global galaxy properties (SFR, gas mass, [C\,\textsc{ii}] luminosity, etc.) were first extracted within a 3D spherical aperture of radius $3\,r_{\rm H}$, where $r_{\rm H}$ is the stellar half-mass radius. The resulting quantities were then projected onto the face-on and edge-on views of the galaxy.
\vspace{-0.25cm}

\subsection{Radiative transfer with Skirt} \label{subsec:Skirt}

In the following, we describe how we generated synthetic observations of \textsc{Vintergatan} using the open-source 3D Monte Carlo radiative transfer code \textsc{Skirt}\footnote{We use version 9 of the code available at \url{https://github.com/Skirt/SKIRT9}} \citep{Camps2015, Camps2020}, to reproduce multi-wavelength observables affected by dust absorption, scattering, and re-emission. To prepare the input medium, we extracted gas cells and star particles from the \textsc{Ramses} snapshots using an octree grid structure. This ensures that the high spatial resolution of the dense regions in \textsc{Vintergatan} was preserved, with the grid refinement levels defined dynamically based on the simulation resolution. We enforced a maximum refinement level such that the finest grid cells in \textsc{Skirt} match the highest-resolution gas cells in the hydrodynamical snapshot. 

For each snapshot, we treated the evolved stellar component (>10 Myr) and the younger stars separately. The former was modelled using the \citet{Bruzual2003} library with a \citet{Chabrier2003} initial mass function. The young star-forming regions was modelled using the \textsc{toddlers} library \citep{Kapoor2023,Kapoor2024}, which provides age- and metallicity-dependent SEDs including nebular and PDR line emission. In \textsc{Skirt}, both stellar components were represented as smoothed particle sources; we adopted the standard cubic-spline smoothing kernel, with the smoothing length of each particle set equal to twice its local cell size (i.e. of order the native RAMSES resolution, $\sim 20$–$100$ pc depending on refinement level), to map the particle luminosities onto the radiative-transfer grid.

The dust distribution was derived directly from the gas cells, from which we extracted positions, metallicities ($Z_{\rm gas}$), masses ($M_{\text{gas}}$), and temperatures ($T_{\text{gas}}$). Gas metallicities $Z_{\rm gas}$ were taken directly from the simulation and correspond to the total metal mass fraction tracked by \textsc{Ramses}, assuming a solar abundance pattern; for stellar particles we additionally computed the metallicity from their Fe and O abundances, $Z_\star \propto 1.06\,{\rm Fe} + 2.09\,{\rm O}$, to match the \textsc{toddlers} input requirements. We computed the dust mass in each cell assuming a constant dust-to-metal ratio, $f_{\text{dust}} = 0.2$, such that $M_{\text{dust}} = f_{\text{dust}}\,Z_{\rm gas}\,M_{\text{gas}}$. The choice of $f_{\rm dust}$ is motivated by the \textsc{Skirt} post-processing of the NewHorizon and NewCluster simulations at $z\simeq 5$ presented in \citet{Ismail2026}, where varying the dust-to-metal ratio by a factor of $\approx 2.5$ only had a modest impact on [C\,\textsc{ii}] luminosities compared to optical tracers, and left the main [C\,\textsc{ii}]–SFR trends unchanged. We restricted the dust distribution to regions where the gas temperature was below $10^6$ K, as dust grains are expected to be destroyed by thermal sputtering in hotter gas, \citep[e.g.][]{Draine1979}. We adopted the \textsc{Themis} dust mix \citep{Jones2017}, which models the physical conditions of dust grains varying between diffuse and dense ISM, with 15 grain size bins for both silicate and hydrocarbon populations. Stochastic heating of dust grains was fully taken into account to accurately model the mid-IR emission.

We accounted for Cosmic Microwave Background (CMB) heating, which is crucial at high redshifts where $T_{\rm CMB}$ becomes comparable to dust temperatures, potentially affecting far-IR luminosity measurements \citep{daCunha2013}. Secondary emission was calculated iteratively until convergence. We followed the photon-statistics prescription of \citet[their Appendix A]{Kapoor2024} to estimate the optimal number of photon packets required to achieve a target signal-to-noise ratio in the synthetic images. In practice, we imposed a fixed physical resolution of 100 pc per pixel at all redshifts and adopted $N_{\rm ph}=10^{10}$ photon packets for all radiative transfer runs, as a compromise between numerical convergence, signal-to-noise ratio, and computational cost. The wavelength grid was discretised into a master logarithmic grid covering 0.01–3000\,$\mu$m, supplemented by high-resolution linear grid centred on the [C\,\textsc{ii}] 158\,$\mu$m emission line. In our set-up we did not explicitly track the carbon density in the hydrodynamical simulation. Instead, the C and C$^+$ abundances and level populations were taken from the pre-computed photoionisation and PDR grids of \textsc{toddlers} at the local gas metallicity, density, and radiation field, assuming a solar abundance pattern scaled by $Z_{\rm gas}$. The [C\,\textsc{ii}] 158\,$\mu$m line emission analysed in this work was taken directly from the resulting \textsc{Skirt} integrated spectra or line cubes.

\begin{figure*}
\centering
\includegraphics{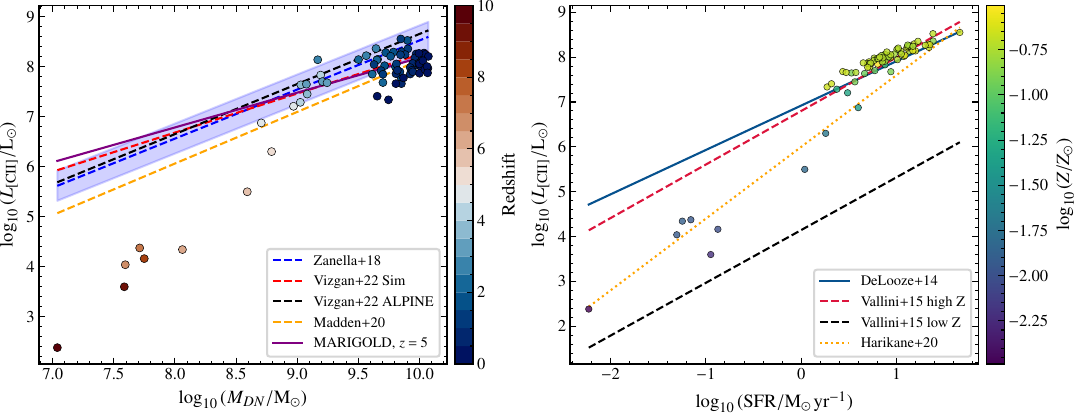}
\caption{(Left panel) Relation between [C\,\textsc{ii}] luminosity and dense neutral gas mass in \textsc{Vintergatan}. Filled circles show the snapshots of \textsc{Vintergatan}, colour-coded by redshift. The blue dashed line and shaded area are the calibration and scatter of \citet{Zanella2018}, derived for main-sequence galaxies with $\log(M_\star/{\rm M_\odot})\approx 10-11$ at $z\approx 2$. The orange dashed line is from \citet{Madden2020}, based on nearby star-forming galaxies and starbursts, and the red and black dashed lines are fits by \citet{Vizgan2022} to their own $z\simeq6$ SIMBA galaxies and ALPINE sources, respectively. The \textsc{Vintergatan} points use the dense neutral gas mass $M_{\rm DN}$, whereas the literature relations are calibrated against molecular gas masses; the comparison therefore also tests the extent to which the dense neutral phase acts as a proxy for the molecular reservoir.
(Right panel) Global $L_{\rm [C\,\textsc{ii}]}$–SFR relation for \textsc{Vintergatan} compared to literature calibrations. Coloured circles show the simulated galaxy, with points colour-coded by global metallicity. The solid blue line denotes the \citet{DeLooze2014} fit to their heterogeneous low-redshift sample, while the black and red dashed lines show the \citet{Vallini2015} metallicity-dependent model evaluated at the highest and $z=7.33$ metallicities reached by \textsc{Vintergatan}. The orange dotted line corresponds to the high-redshift fit of \citet{Harikane2020} for [C\,\textsc{ii}]-deficient galaxies at $z\approx 6-9$.}
\label{fig:global_relations}
\end{figure*}

\vspace{-0.25cm}
\subsection{Extraction of the [CII] line measurements from integrated and resolved SEDs}

For each snapshot, we derived integrated [C\,\textsc{ii}] properties directly from the spatially integrated SEDs produced by \textsc{Skirt}. For the [C\,\textsc{ii}] 158\,$\mu$m line, we converted the rest-frame line wavelength to the observed frame using the snapshot redshift and define a velocity window, typically $\pm 2000$ km s$^{-1}$, to enclose the line profile and neighbouring continuum.

Within this spectral window, we estimated the underlying continuum by fitting a linear function to line-free channels on both sides of the line centre, excluding the region where the line emission dominates. After subtracting this continuum model, we fited a single Gaussian profile to the residual line spectrum using standard least-squares minimisation to obtain the line centroid, width, and peak amplitude. The integrated line flux was then computed analytically from the best-fitting Gaussian parameters and converted to a line luminosity as
\begin{equation}\label{eq:L_cii}
    L_{\rm [C\,\textsc{ii}]} = 4\pi D_{\rm L}^2 
    \int_{\Delta v} \left[ F_\nu(v) - F_{\nu,{\rm cont}}(v) \right] \, {\rm d}v,
\end{equation}
where $D_{\rm L}$ is the luminosity distance at the snapshot redshift, $F_\nu(v)$ is the observed-frame flux density as a function of velocity relative to the [C\,\textsc{ii}] line centre, and $F_{\nu,{\rm cont}}(v)$ is the local continuum level obtained from the linear continuum fit. The integral was computed over the chosen velocity window $\Delta v$ around the line. These global measurements provided a consistent set of [C\,\textsc{ii}] luminosities and line widths for each snapshot and orientation and were used for all analyses that rely on galaxy-integrated quantities.

In addition, we applied the same continuum subtraction and single-Gaussian fitting procedure to [C\,\textsc{ii}] spectra extracted in each pixel of the 3D cubes, obtaining a resolved [C\,\textsc{ii}] line-flux map. After rejecting non-convergent or unphysical fits (e.g. negative amplitudes or extremely broad widths), we converted this flux map into a [C\,\textsc{ii}] luminosity map using the luminosity distance at the snapshot redshift, and then into a [C\,\textsc{ii}] surface-brightness map by dividing by the physical area of each pixel. These $\Sigma_{\rm [C\,\textsc{ii}]}$ maps formed the basis of our spatially resolved analysis and are compared to the corresponding $\Sigma_{\rm SFR}$ maps.

\subsection{Projection and aperture size}

All the steps described above were applied independently to different viewing angles of the same galaxy. For each snapshot, we computed the total stellar angular-momentum vector and defined a face-on view by aligning the simulation $z$-axis with this vector, and an edge-on view by rotating the line of sight by $90^\circ$. Because the [C\,\textsc{ii}] 158\,$\mu$m line lies in the far-infrared regime, dust attenuation and viewing-angle effects have a negligible impact on the total [C\,\textsc{ii}] luminosity. We have further verified that restricting the 3D aperture from $3\,r_{\rm H}$ to $1\,r_{\rm H}$ only weakly changes the inferred $\alpha_{\rm [C\,\textsc{ii}]}$, owing to the modest variations in enclosed dense neutral fraction and metallicity over this range of radii. We therefore adopted the face-on view and a spherical aperture of radius $3\,r_{\rm H}$ as our fiducial configuration, and relegated a detailed comparison of different orientations, apertures, star formation timescale, and transparent versus the full radiative-transfer [C\,\textsc{ii}] luminosities to Appendix~\ref{apdx:left_out}, where we showed that these choices do not qualitatively alter the [C\,\textsc{ii}] luminosity–SFR scaling relation.

\section{Results}\label{sec:results}
\subsection{Global scaling relations and consistency with literature} \label{subsec:global_scaling}

To benchmark our simulation against observational constraints, we compare the total SFR and dense neutral gas mass ($M_{\rm DN}$) to the [C\,\textsc{ii}] luminosity ($L_{\rm [C\,\textsc{ii}]}$), integrated within three stellar half-mass radii. In addition, we consider the ratio between [C\,\textsc{ii}] and infrared luminosities (integrated between $8$–$1000~\mu{\rm m}$), $L_{\rm [C\,\textsc{ii}]}/L_{\rm IR}$.

The left panel of Fig.~\ref{fig:global_relations} presents $L_{\rm [C\,\textsc{ii}]}$ as a function of $M_{\rm DN}$ at different redshifts. At $z \lesssim 5$, \textsc{Vintergatan} lies close to the empirical, observation-based $L_{\rm [C\,\textsc{ii}]}$–$M_{\rm mol}$ relation of \citet{Zanella2018}, derived for main-sequence galaxies at $z\approx 2$ with $\log(M_\star/{\rm M_\odot})\approx 10-11$. In their calibration, $M_{\rm mol}$ is inferred from CO(1–0) and dust continuum measurements, converted to molecular gas masses using standard, metallicity-dependent CO–to–H$_2$ and dust–to–gas conversion factors. Around $z\simeq2$, our points overlap their relation particularly well, while at lower redshift they remain within the scatter of the low- and intermediate-redshift samples compiled by \citet{Madden2020} and \citet{Vizgan2022}. These studies provide $L_{\rm [C\,\textsc{ii}]}$–$M_{\rm mol}$ and $L_{\rm [C\,\textsc{ii}]}/L_{\rm IR}$ benchmarks spanning nearby star-forming galaxies and higher-redshift systems, against which \textsc{Vintergatan} reproduces the same [C\,\textsc{ii}] output of chemically evolved, main-sequence galaxies.
A fit of the form
\begin{equation}
    \log_{10} L_{\rm [C\,\textsc{ii}]} = a\,\log_{10} M_{\rm DN} + b
\end{equation}
to the snapshots $0 < z < 6$ yields $a_{\rm low} \simeq 0.94$ and an intrinsic scatter of $\approx 0.2-0.3\,\rm{dex}$, consistent with a nearly constant [C\,\textsc{ii}]–to–dense neutral–gas conversion factor,
\begin{equation}\label{eq:alpha_glob}
    \alpha_{\rm [C\,\textsc{ii}]} \equiv \frac{M_{\rm DN}}{L_{\rm [C\,\textsc{ii}]}},
\end{equation}
in this regime. At earlier times ($6 \lesssim z \lesssim 10$) however, the simulated points lie systematically below the observed $L_{\rm [C\,\textsc{ii}]}$–$M_{\rm mol}$ relations and define a steeper sequence than the approximately linear trends implied by these calibrations: the best-fitting relation has a slope of $a_{\rm high} \simeq 2.41$, with an intrinsic scatter similar to the low-redshift branch. This behaviour indicates that, along the \textsc{Vintergatan} track, the [C\,\textsc{ii}] luminosity increases more rapidly than the dense neutral gas mass during the early assembly of the galaxy.

\begin{figure}
\centering
\includegraphics{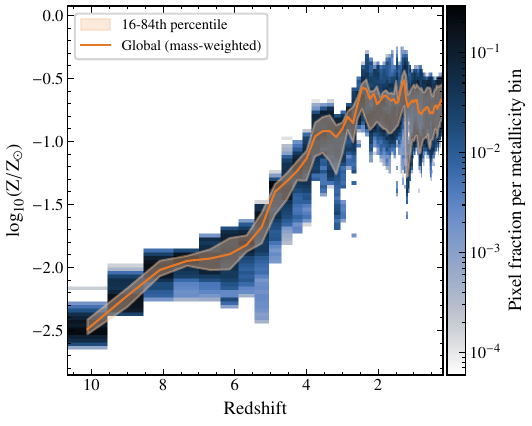}
\caption{Redshift evolution of the gas metallicity distribution in \textsc{Vintergatan}. The colour map shows, at each redshift, the fraction of pixels per metallicity bin, while the orange line and shaded band give the mass-weighted average metallicity and the 16th–84th percentiles of the pixel distribution.}
\label{fig:Z_vs_z}
\end{figure}

In this early phase, the simulated $L_{\rm [C\,\textsc{ii}]}/L_{\rm IR}$ ratios are reduced by up to two orders of magnitude relative to the typical late-time plateau around $10^{-3}$. These values are therefore lower than those measured in the most [C\,\textsc{ii}]-deficient local ULIRGs and in current high-$z$ [C\,\textsc{ii}]-bright samples, which typically show $L_{\rm [C\,\textsc{ii}]}/L_{\rm IR} \gtrsim 3\times10^{-4}$ \citep[e.g.][]{Luhman2003,Harikane2020}. In our simulation this extreme [C\,\textsc{ii}] deficit appears when the global gas metallicity is very low (Fig.~\ref{fig:Z_vs_z}), and coincides with the regime where the global $L_{\rm [C\,\textsc{ii}]}$–$M_{\rm DN}$ and $L_{\rm [C\,\textsc{ii}]}$–SFR relations are steeper than the nearly linear calibrations derived in the local Universe. Together, these trends show that the [C\,\textsc{ii}]–to–dense neutral–gas conversion factor, $\alpha_{\rm [C\,\textsc{ii}]}$, varies by several orders of magnitude in \textsc{Vintergatan}, and cannot be represented by a single constant value.

The same deviation from linearity is obtained when replacing dense neutral gas by SFR, as shown in the right panel of Fig.~\ref{fig:global_relations}. At $z \gtrsim 5$, when \textsc{Vintergatan} is still a low-mass, metal-poor system, with $M_\star \approx 3\times10^{7}\,{\rm M_\odot}$ and $\log_{10} (Z_{\rm gas}/{\rm Z_\odot}) \lesssim -1.3$, the galaxy follows a very steep sequence: $L_{\rm [C\,\textsc{ii}]}$ rises by nearly five orders of magnitude while the SFR increases by only $\approx 3$ dex. This behaviour reflects the rapid global chemical enrichment shown in Fig.~\ref{fig:Z_vs_z}: the mass-weighted metallicity starts from very metal-poor regimes ($\log_{10} (Z_{\rm gas}/{\rm Z_\odot}) \simeq -2.5$ at $z \approx 10$) and only exceeds $\log_{10} (Z_{\rm gas}/{\rm Z_\odot}) \simeq -1.3$ at $z \approx 5$, where the change of regime appears in both panels of Fig.~\ref{fig:global_relations}. The resolved metallicity distribution shown by the coloured density map in Fig.~\ref{fig:Z_vs_z} closely tracks this global trend, with the 16th–84th percentiles forming a narrow band of $\approx 0.2~{\rm dex}$, indicating that the ISM is well-mixed.
We note that recent cosmological simulations that explicitly follow molecular gas and [C\,\textsc{ii}] emission, such as \textsc{ColdSIM} \citep{Casavecchia2024} and MARIGOLD \citep{Khatri2025}, predict $L_{\rm [C\,\textsc{ii}]}$–SFR and $L_{\rm [C\,\textsc{ii}]}$–$M_{\rm mol}$ relations with slopes and normalisations consistent with the other literature fits used here. At fixed SFR or gas mass, these relations yield [C\,\textsc{ii}] luminosities systematically higher than for \textsc{Vintergatan}, by factors of a few. This reflects the fact that they are calibrated on more massive and more metal-rich galaxy populations; the resulting offset therefore primarily traces differences in galaxy mass and metallicity rather than the details of our phase separation.

Taken together, these results suggest that, at early times when the global gas metallicity is still low, the evolution of the [C\,\textsc{ii}] luminosity is primarily driven by the changing ISM conditions and, in particular, by the enrichment of metals. Studies based on FIRE-2 zoom-in simulations at $z\gtrsim5$ likewise find an approximately linear $L_{\rm [C\,\textsc{ii}]}$–SFR relation at SFRs above a few ${\rm M_\odot\,yr^{-1}}$ that matches current ALMA samples \citep{Liu2026}, but do not probe the low-SFR regime ($\mathrm{SFR}\lesssim 1\,{\rm M_\odot\,yr^{-1}}$) where our \textsc{Vintergatan} results reveal a steeper, strongly evolving $L_{\rm [C\,\textsc{ii}]}$–SFR relation. Once the gas is sufficiently enriched, with $\log_{10} (Z_{\rm gas}/{\rm Z_\odot})$ settling on a plateau around $-1$ to $-0.5$ after $z \approx 4$, the [C\,\textsc{ii}]–SFR and [C\,\textsc{ii}]–$M_{\rm DN}$ relations approach the widely used, nearly linear relation discussed above for $z \lesssim 5$.

To contextualise this evolution, we compare our results with literature calibrations. In the right panel of Fig.~\ref{fig:global_relations}, we include the 'entire sample' $L_{\rm [C\,\textsc{ii}]}$–SFR relation of \citet{DeLooze2014}, derived from a heterogeneous set of over $\sim 500$ low-redshift systems ranging from dwarf galaxies to ULIRGs. We note that using the other \citet{DeLooze2014} relations restricted to specific subsamples (e.g. starbursts or metal-poor dwarfs) does not alter the qualitative comparison: the low-metallicity, i.e. high-$z$, \textsc{Vintergatan} points consistently fall below these local observational benchmarks.

We also compare with the metallicity-dependent theoretical model of \citet{Vallini2015}, showing tracks evaluated at two fixed metallicities that roughly bracket the \textsc{Vintergatan} data: a low-metallicity benchmark ($\log_{10} (Z_{\rm gas}/{\rm Z_\odot}) = -1.95$) and an enriched benchmark ($\log_{10} (Z_{\rm gas}/{\rm Z_\odot}) = -0.5$). In the low-$Z$, high-redshift regime, the \textsc{Vintergatan} points lie systematically below the \citet{DeLooze2014} relation and the high-metallicity \citet{Vallini2015} track. However, the apparent offset from the \citet{Vallini2015} relation is largely driven by the low metallicity of \textsc{Vintergatan} at these epochs. If instead we evaluate the \citet{Vallini2015} prescription using, for each snapshot, the 10 Myr-averaged SFR and the gas metallicity measured in \textsc{Vintergatan}, the resulting model still lies systematically below the [C\,\textsc{ii}] luminosities obtained from our radiative transfer post-processing with \textsc{Skirt}. This demonstrates that, for fixed physical conditions (SFR, $Z_{\rm gas}$), the simulated [C\,\textsc{ii}] emission is brighter than predicted by the analytical model. We caution, however, that this discrepancy likely arises from the calibration of the \citet{Vallini2015} relation. Their fitting formula is derived for metallicities $Z_{\rm gas}\gtrsim 0.05\,{\rm Z_\odot}$, and applying it to the extremely metal-poor regime of \textsc{Vintergatan} at high redshift requires extrapolating well outside its domain of validity.

Instead, the global \textsc{Vintergatan} measurements at $z \gtrsim 5$ closely follow the empirical high-$z$ $L_{\rm [C\,\textsc{ii}]}$–SFR relation of \citet{Harikane2020}, derived for nine luminous $z \approx 6-9$ galaxies that exhibit a pronounced [C\,\textsc{ii}] deficit. We note that these objects are predominantly Ly$\alpha$ emitters with relatively low metallicities ($Z_{\rm gas}\approx 0.01 {-} 0.1\,Z_\odot$), whereas other high-redshift [C\,\textsc{ii}] samples used in the literature are typically massive, metal-enriched systems in the regime where low-$z$, nearly linear relations apply. The agreement with the \citet{Harikane2020} relation indicates that, in this low-metallicity phase, \textsc{Vintergatan} occupies the same $L_{\rm [C\,\textsc{ii}]}$–SFR locus as [C\,\textsc{ii}]-deficient high-redshift galaxies, even though it is less massive. As the galaxy becomes more metal-rich ($\log_{10} (Z_{\rm gas}/{\rm Z_\odot}) \gtrsim -1.3$), a second, shallower sequence emerges: the $L_{\rm [C\,\textsc{ii}]}$–SFR relation approaches a nearly linear behaviour with reduced scatter, and $L_{\rm [C\,\textsc{ii}]}$ scales almost proportionally with SFR. In this enriched regime, the \textsc{Vintergatan} points are consistent with the \citet{DeLooze2014} calibration and rise to meet (and occasionally exceed) the $\log_{10} (Z_{\rm gas}/{\rm Z_\odot}) = -0.5$ \citet{Vallini2015} track, indicating that the simulated system recovers [C\,\textsc{ii}] luminosities comparable to those of low-$z$, chemically evolved star-forming galaxies once its ISM has reached moderate metallicities.

These results demonstrate that \textsc{Vintergatan}'s [C\,\textsc{ii}] scaling relations evolve strongly with cosmic time. Local-Universe calibrations apply once the ISM is enriched with metals, but a different prescription is required at high redshift, with the very-early phase ($L_{\rm [C\,\textsc{ii}]}/L_{\rm IR} \approx 10^{-5}$) probing conditions more extreme than those reached in current high-redshift [C\,\textsc{ii}] samples.

\subsection{Global and resolved \texorpdfstring{$\alpha_{[\mathrm{CII}]}$}{alpha\_[CII]} conversion factor in context} \label{subsec:resolved_alpha}

The strong redshift evolution of the [C\,\textsc{ii}]–to–dense neutral gas relation in \textsc{Vintergatan}, and its tension with existing constraints on a nearly constant $\alpha_{\rm [C\,\textsc{ii}]}$ at low and intermediate redshift \citep[e.g.][]{Zanella2018,Rowland2024}, shows that a single, universal conversion factor cannot describe the ISM across cosmic time. To understand whether the observed [C\,\textsc{ii}] scaling and deficits arise from galaxy-wide ISM properties or are driven by localised, high-density star-forming regions, we compare global and spatially resolved estimates of the [C\,\textsc{ii}]-to-dense neutral gas conversion factor, $\alpha_{\rm [C\,\textsc{ii}]}$.

\begin{figure}
\centering
\includegraphics{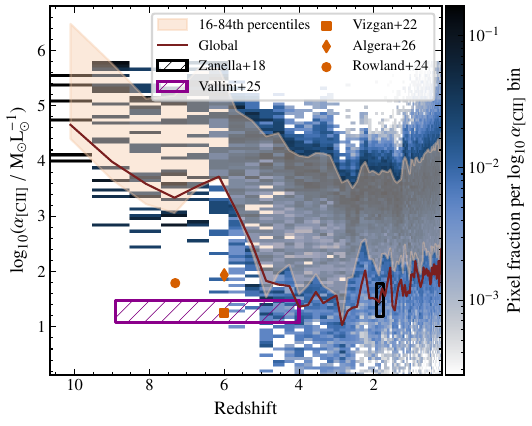}   
\caption{Redshift evolution of the [C\,\textsc{ii}]–to–dense neutral-gas conversion factor $\alpha_{\rm [C\,\textsc{ii}]}$. The blue colour map shows the distribution of pixel-wise $\alpha_{\rm [C\,\textsc{ii}]}$ values, the dark red line traces the global conversion factor, and the orange shaded region marks the 16th–84th percentile range of the resolved values. Black and purple hatched rectangles indicate the ranges from the \citet{Zanella2018} calibration and the \textsc{SERRA} simulations of \citet{Vallini2025}, respectively, while individual orange symbols show measurements for a low-mass $z\approx 6$ galaxy \citep[][square]{Vizgan2022}, its revised value from \citet[][diamond]{Algera2026}, and the $z=7.31$ disc of \citet[][circle]{Rowland2024}.}
\label{fig:alpha_evolution}
\end{figure}

Figure~\ref{fig:alpha_evolution} presents the redshift evolution of $\alpha_{\rm [C\,\textsc{ii}]}$, comparing the global conversion factor, obtained from the ratio of the total dense neutral gas mass to the total [C\,\textsc{ii}] luminosity in each snapshot, to the distribution of local conversion factors measured on a pixel–by–pixel basis. For each pixel we define
\begin{equation}
    \alpha_{\rm [C\,\textsc{ii}]}(x,y) \equiv \frac{\Sigma_{\rm DN}(x,y)}{\Sigma_{\rm [C\,\textsc{ii}]}(x,y)}, \label{eq:alpha_reso}
\end{equation}
where $\Sigma_{\rm DN}$ and $\Sigma_{\rm [C\,\textsc{ii}]}$ are the dense neutral gas and [C\,\textsc{ii}] surface densities, respectively, expressed in ${\rm M_\odot\,kpc^{-2}}$ and ${\rm L_\odot\,kpc^{-2}}$. We work with surface densities for the resolved $\alpha_{\rm [C\,\textsc{ii}]}$ because the \textsc{Skirt} output provides [C\,\textsc{ii}] surface-brightness maps, and most observational applications likewise infer $\alpha_{\rm [C\,\textsc{ii}]}$ from projected quantities (luminosities and areas) rather than from true 3D volume densities. Each resolved pixel corresponds to an element of $100\,{\rm pc} \times 100\,{\rm pc}$.

We overplot literature estimates derived for different galaxy populations: the widely used calibration of \citet{Zanella2018} for star-forming galaxies at $1.73<z<1.94$; \citet{Rowland2024} for a dynamically cold disc massive galaxy at $z=7.31$; \citet{Vizgan2022} for a low-mass $z=6$ system with $\alpha_{\rm [C\,\textsc{ii}]}=18\,{\rm M_\odot\,L_\odot^{-1}}$; the revised value $\alpha_{\rm [C\,\textsc{ii}]}=87\,{\rm M_\odot\,L_\odot^{-1}}$ for the same source from \citet{Algera2026}; and the mean value inferred for 98 galaxies with $M_\star >10^9~{\rm M_\odot}$ at $4<z<8.9$ from the \textsc{SERRA} simulations analysed by \citet{Vallini2025}. The \textsc{SERRA} galaxies broadly overlap with the mass range reached by \textsc{Vintergatan} at $z \lesssim 5$, but at fixed redshift they typically probe systems that are more massive and more metal-rich; the comparison should therefore be viewed as indicative rather than direct. All of these benchmarks correspond to galaxy-integrated measurements or simulated analogues, obtained by converting observed [C\,\textsc{ii}] luminosities and independent estimates of molecular gas mass into a global $\alpha_{\rm [C\,\textsc{ii}]}$.

At $z \leq 5$, \textsc{Vintergatan} yields $\alpha_{\rm [C\,\textsc{ii}]}\approx 10$–$10^2\,{\rm M_\odot\,L_\odot^{-1}}$, and the global curve lies within the scatter of the \citet{Zanella2018} calibration and of the lowest-redshift part of the \textsc{SERRA} envelope, indicating broad consistency with current empirical and theoretical constraints once the ISM is enriched. In this regime the galaxy has reached $M_\star \gtrsim 10^9~{\rm M_\odot}$ and $\log_{10}(Z_{\rm gas}/{\rm Z_\odot}) \gtrsim -1.3$, and our results support the use of a nearly constant $\alpha_{\rm [C\,\textsc{ii}]}$ similar to those adopted in $z\approx 0$ to $z \approx 5$ studies. At earlier times ($z > 5$), the simulation predicts much larger values of $\alpha_{\rm [C\,\textsc{ii}]}$, reaching $\gtrsim 10^3\,{\rm M_\odot\,L_\odot^{-1}}$ by $z \approx 6$–8, whereas individual galaxies such as the \citet{Vizgan2022,Rowland2024,Algera2026} systems and most of the \textsc{SERRA} population remain in the $\alpha_{\rm [C\,\textsc{ii}]}\approx 10$–$10^2~{\rm M_\odot\,L_\odot^{-1}}$ regime. We therefore infer a systematic mismatch of one to two orders of magnitude between our results and the presently observed [C\,\textsc{ii}]-bright galaxy population at high redshift. This reflects the fact that \textsc{Vintergatan} in this phase is substantially less massive and more metal-poor than the galaxies used to calibrate nearly constant $\alpha_{\rm [C\,\textsc{ii}]}$ relations, and therefore probes a region of parameter space where [C\,\textsc{ii}] is intrinsically faint and a larger conversion factor is required. At these very early times, \textsc{Vintergatan} exhibits both low dense neutral gas masses and extremely low gas-phase metallicities, yet $\alpha_{\rm [C\,\textsc{ii}]}$ reaches values $\gtrsim 10^3\,{\rm M_\odot\,L_\odot^{-1}}$. This indicates that the [C\,\textsc{ii}] luminosity becomes increasingly suppressed relative to the dense neutral gas content as one moves to higher redshift and lower mass. Such behaviour is qualitatively consistent with increasingly metal-loaded outflows at low stellar mass, which can remove freshly produced metals more rapidly than the dense neutral reservoir is depleted \citep[e.g.][]{Peeples2014,Ma2016}. This picture is supported by previous \textsc{Vintergatan} analyses, which find that outflows are systematically more metal-rich than inflows by $\sim 0.5$ dex throughout the galaxy’s assembly, including during the merger-dominated growth phase \citep{Renaud2021a}. By lowering the carbon abundance in the cold ISM, these winds reduce the [C\,\textsc{ii}] emissivity per unit dense neutral mass even when $M_{\rm DN}$ remains non-negligible.

The relation between the global $\alpha_{\rm [C\,\textsc{ii}]}$ and the median of the resolved pixel distribution provides direct insight into how representative a single, galaxy-integrated conversion factor is of the underlying local conditions. At high redshift ($z \gtrsim 6$), these two quantities decrease together by nearly three orders of magnitude and typically agree within a factor of a few, indicating that the strong evolution of the global conversion factor reflects a coherent shift of the full distribution of local $\alpha_{\rm [C\,\textsc{ii}]}$ values with redshift rather than being driven by a few extreme regions. At lower redshifts, we instead find a clear offset between the galaxy-integrated $\alpha_{\rm [C\,\textsc{ii}]}$ and the median resolved value. In this regime, the global $\alpha_{\rm [C\,\textsc{ii}]}$ is dominated by a relatively small number of bright [C\,\textsc{ii}]-emitting regions with moderate $\alpha_{\rm [C\,\textsc{ii}]}$, while many fainter pixels, comparatively poor in dense neutral gas, populate the high-$\alpha_{\rm [C\,\textsc{ii}]}$ tail of the resolved distribution. The resulting difference between the global and median of the resolved $\alpha_{\rm [C\,\textsc{ii}]}$ therefore traces a genuine change in how [C\,\textsc{ii}] and dense neutral gas are distributed within the galaxy. For completeness, we also construct an analogous version of Fig.~\ref{fig:alpha_evolution} as a function of the global, mass-weighted gas metallicity (Fig.~\ref{fig:alpha_vs_Z}). This metallicity-based view shows the same behaviour, indicating that the apparent redshift dependence of $\alpha_{\rm [C\,\textsc{ii}]}$ is primarily driven by the build-up of metals rather than by cosmic time itself. The main takeaway from Figs.~\ref{fig:alpha_evolution} and \ref{fig:alpha_vs_Z} is that both global and resolved $\alpha_{\rm [C\,\textsc{ii}]}$ evolve strongly along the \textsc{Vintergatan} track, with their behaviour primarily governed by the evolving metallicity and ISM structure; even in the enriched regime, a single, universal conversion factor cannot capture the full range of local conditions sampled within the galaxy.

\subsection{Prescriptions for \texorpdfstring{$\alpha_{\rm [CII]}$}{alpha\_[CII]} in VINTERGATAN}
\label{subsec:prescriptions_relations}

\textsc{Vintergatan} differs markedly from the massive, [C\,\textsc{ii}]-bright systems on which existing calibrations are based. Thus, we propose tailored prescriptions for $\alpha_{\rm [C\,\textsc{ii}]}$. Applying, for example, the metallicity-dependent formula of \citet[][their Eq.~10]{Vallini2025} to \textsc{Vintergatan} requires extrapolating far below its calibrated metallicity range and yields conversion factors that recover at most $\sim 60 \%$ of the true dense neutral mass once the galaxy has enriched, and essentially none of it at earlier, very metal-poor epochs. This demonstrates that current relations do not capture the strong increase of $\alpha_{\rm [C\,\textsc{ii}]}$ in the low-metallicity regime probed by \textsc{Vintergatan}, and motivates constructing practical prescriptions directly from our simulation.

We adopt simple log–log forms for these prescriptions primarily for practical reasons. Most observational [C\,\textsc{ii}]–SFR and [C\,\textsc{ii}]–gas calibrations are expressed as power laws in luminosity, surface density, and metallicity, and are straightforward to fit and compare. Our functional forms are therefore deliberately close to those used in the existing literature, allowing our simulation-based calibrations to be implemented and contrasted directly with observational ones.

Guided by these considerations, we fitted log–log relations for both the global and local definitions of $\alpha_{\rm [C\,\textsc{ii}]}$ (Eqs.~\ref{eq:alpha_glob} and \ref{eq:alpha_reso}). For the global measurements we write
\begin{align}
\log_{10}\alpha_{\rm [C\,\textsc{ii}]}^{\rm glob} &=
A_{\rm glob}\,\log_{10}L_{\rm [C\,\textsc{ii}]} + B_{\rm glob}\,\log_{10}(Z_{\rm gas}/Z_\odot) + C_{\rm glob}, \label{eq:glob_L_Z} \\
\log_{10}\alpha_{\rm [C\,\textsc{ii}]}^{\rm glob} &=
B_{\rm glob}\,\log_{10}(Z_{\rm gas}/Z_\odot)+ C_{\rm glob}, \label{eq:glob_Z_only}
\end{align}
where $L_{\rm [C\,\textsc{ii}]}$ is the integrated [C\,\textsc{ii}] luminosity of the galaxy and $Z_{\rm gas}$ denotes the global, gas-mass-weighted metallicity.

For the resolved measurements we adopt analogous forms:
\begin{align}
\log_{10}\alpha_{\rm [C\,\textsc{ii}]} &=
A_{\rm res}\,\log_{10}\Sigma_{\rm [C\,\textsc{ii}]} +
B_{\rm res}\,\log_{10}(Z_{\rm gas}/Z_\odot) + C_{\rm res}, \label{eq:reso_Sigma_Z}\\
\log_{10}\alpha_{\rm [C\,\textsc{ii}]} &=
A_{\rm res}\,\log_{10}\Sigma_{\rm [C\,\textsc{ii}]} + C_{\rm res}, \label{eq:reso_Sigma_only}
\end{align}
where $\Sigma_{\rm [C\,\textsc{ii}]}$ is the local [C\,\textsc{ii}] surface brightness and $Z_{\rm gas}$ the local gas metallicity.

\begin{table}[ht]
    \centering
    \caption{Fit parameters for global and resolved measurements.}
    \begin{tabular}{cccccc}
         \hline\hline
         Method & Range & A & B & C & $\sigma_i$\\
         \hline
         \multirow{3}{*}{\shortstack[c]{Global\\($L_{\rm [C\,\textsc{ii}]} + Z_{\rm gas})$}} 
           & All            & -0.757 &  0.837 & 8.455 & 0.226 \\
           & $z\geq4.00$    & -0.405 & -0.797 & 3.599 & 0.146 \\
           & $z\geq5.25$    & -0.426 & -0.535 & 4.318 & 0.164 \\\hline
        \multirow{3}{*}{\shortstack[c]{Global\\(Z only)}} 
            & All           & --     & -1.225 &  0.897 & 0.373 \\
            & $z\geq4.00$   & --     & -2.514 & -1.498 & 0.209 \\
            & $z\geq5.25$   & --     & -2.358 & -1.161 & 0.250 \\\hline      
         \multirow{3}{*}{\shortstack[c]{Resolved\\($\rm \Sigma_{[C\,\textsc{ii}]} + Z_{\rm gas})$}}
            & All           & -0.657 &  0.093 & 6.413 & 0.561 \\
            & $z\geq4.00$   & -0.685 & -0.371 & 5.971 & 0.549 \\
            & $z\geq5.25$   & -0.657 & -0.390 & 5.786 & 0.581 \\\hline
        \multirow{3}{*}{\shortstack[c]{Resolved\\($\Sigma_{\rm[C\,\textsc{ii}]}$ only)}}
            & All           & -0.653 & --     & 6.320 & 0.562 \\
            & $z\geq4.00$   & -0.720 & --     & 6.661 & 0.559 \\
            & $z\geq5.25$   & -0.675 & --     & 6.631 & 0.582 \\\hline

    \end{tabular}
    \label{tab:fits}
    \tablefoot{
        Parameters of fits to multiple snapshots, for global and resolved measurements, with and without gas metallicity as an additional parameter. The corresponding fitting equations are given in Eqs.~\ref{eq:glob_L_Z}--\ref{eq:reso_Sigma_Z}.
    }
\end{table}

\begin{figure}
\centering
\includegraphics{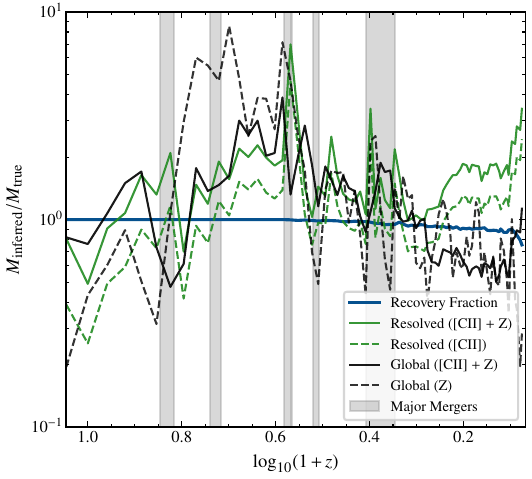}
\caption{Recovery fraction of the total dense neutral gas mass inferred from different [C\,\textsc{ii}]–based estimators as a function of redshift. The blue line shows the recovery fraction when the true dense neutral surface density is summed only over [C\,\textsc{ii}]–emitting pixels. The green solid and dashed lines show the results from the resolved relations including both $\Sigma_{\rm[C\,\textsc{ii}]}$ and metallicity, and $\Sigma_{\rm[C\,\textsc{ii}]}$ alone, respectively. The black dashed and solid lines show the results from the global prescriptions using, respectively, metallicity only and $L_{\rm [C\,\textsc{ii}]}+Z_{\rm gas}$. Values close to unity indicate accurate recovery, grey shaded bands denote major merger events identified in \cite{Renaud2021a}.}
\label{fig:mass_inference_bias}
\end{figure}

For the resolved fits, we restricted the calibration to pixels in which the [C\,\textsc{ii}] surface brightness, dense neutral gas surface density, and metallicity are all positive, ensuring that $\alpha_{\rm [C\,\textsc{ii}]}$ is well defined. Pixels with detectable [C\,\textsc{ii}] but no dense neutral gas, or with dense neutral gas but no [C\,\textsc{ii}], were excludeded from the fit and enter only through the reference 'true' dense neutral mass against which we compare the inferred values in Fig.~\ref{fig:mass_inference_bias}. In practice, $\sim 67\%$ of valid pixels have $\Sigma_{\rm [C\,\textsc{ii}]} > 0$ but $\Sigma_{\rm DN} = 0$ according to our phase definition, yet these pixels contribute only $\sim 1\%$ of the total [C\,\textsc{ii}] luminosity, and we find no pixels with $\Sigma_{\rm DN} > 0$ and undetected [C\,\textsc{ii}] at this resolution. This shows that essentially all of the [C\,\textsc{ii}] emission and all of the dense neutral mass arise from regions where both components are present, and that the resolved fits are dominated by overlapping [C\,\textsc{ii}]-bright, dense neutral-gas-rich pixels.

In a second step, we apply the resolved prescriptions directly to the [C\,\textsc{ii}] maps, as would be done in an observational analysis, to test how well they recover the total dense neutral gas mass from [C\,\textsc{ii}] alone. We evaluate the best-fitting local relation $\alpha_{\rm [C\,\textsc{ii}]}(\Sigma_{\rm [C\,\textsc{ii}]}, Z_{\rm gas})$ for every [C\,\textsc{ii}]-emitting pixel and convert the [C\,\textsc{ii}] surface brightness into an inferred dense neutral surface density, irrespective of whether the pixel contains dense neutral gas in the simulation. Summing this inferred dense neutral map over all [C\,\textsc{ii}]-bright pixels then yields a predicted total dense neutral mass, which we compare to the true value in Fig.~\ref{fig:mass_inference_bias}. This procedure allows us to quantify whether the prescriptions calibrated on overlapping [C\,\textsc{ii}]-bright, dense neutral-gas-rich regions systematically over- or underestimate the dense neutral reservoir when applied to realistic [C\,\textsc{ii}] maps.

For each choice of functional form (global or resolved, with or without metallicity dependence) we infer the slope(s), intercept, and intrinsic scatter $(A,B,C,\sigma_i)$ using an MCMC sampler that explores the posterior distribution of the parameters given all available snapshots in the selected redshift range. We model the likelihood as a Gaussian in $\log_{10}\alpha_{\rm [C\,\textsc{ii}]}$ around the proposed relation with intrinsic scatter $\sigma_i$, and sample the posterior for $(A,B,C,\sigma_i)$ with 32 walkers evolved for 3000 steps using \textsc{emcee} \citep{foreman-mackey2013}, discarding the initial burn-in. In addition to a fit over the full redshift range, we also derive fits restricted to an enriched regime at $z \geq 4.0$, where the ISM metallicity has largely settled, and to a primordial regime at $z \geq 5.25$, which isolates the extremely metal-poor, rapidly evolving phase identified in Sect.~\ref{subsec:global_scaling}. These restricted fits, summarised in Table~\ref{tab:fits} and discussed in Appendix~\ref{apdx:bias}, are used to test whether calibrating $\alpha_{\rm [C\,\textsc{ii}]}$ directly in the relevant physical regime improves the recovery of the total dense neutral gas mass.

Figure~\ref{fig:mass_inference_bias} summarises how these prescriptions perform when used to recover the total dense neutral gas mass. As a check, the blue curve shows the 'recovered mass fraction', defined by summing the true dense neutral surface density only over [C\,\textsc{ii}]–emitting pixels. This quantity stays very close to unity at all redshifts, with deviations of at most $\sim 10 \%$ at late times, demonstrating that [C\,\textsc{ii}] emission traces almost the entire dense neutral gas reservoir at all redshifts. The larger deviations from unity discussed below therefore arise from the conversion factor itself, rather than from a substantial [C\,\textsc{ii}]-dark dense neutral-gas component.

The solid and dashed green curves show the bias in the inferred mass when the resolved relations of Eqs.~\eqref{eq:reso_Sigma_Z} and \eqref{eq:reso_Sigma_only} are applied pixel by pixel and integrated over the galaxy, using Eq.~\eqref{eq:alpha_reso} to convert $\Sigma_{\rm [C\,\textsc{ii}]}$ (and, where relevant, $Z_{\rm gas}$) into a dense neutral surface density in each pixel. Both prescriptions behave similarly and, once evaluated on all [C\,\textsc{ii}]-emitting pixels, recover the total dense neutral mass to within factors of order unity over most of the redshift range, with fluctuations at the tens-of-per-cent level that correlate with the merger epochs. Including metallicity in the resolved fit yields a modest but systematic improvement at very high redshift, reducing the gap between the inferred mass and true mass by $\sim 10 \%$ between $z\simeq 10$ and $z\simeq 6$. This indicates that $\Sigma_{\rm [C\,\textsc{ii}]}$ and $Z_{\rm gas}$ do capture a large part of the local dependence of $\alpha_{\rm [C\,\textsc{ii}]}$ in \textsc{Vintergatan}, but that additional local properties (e.g. gas density, radiation field, or the relative contributions of dense neutral, diffuse neutral, and ionised gas) are still required to stabilise the resolved conversion factor fully, especially during mergers.

The global prescriptions vary differently. The black solid curve corresponds to the metallicity–only model, while the black dashed curve shows our bivariate luminosity–metallicity fit (Eq.~\ref{eq:glob_L_Z}). The $Z_{\rm gas}$–only relation performs poorly at all times: it underestimates the dense neutral mass at the highest redshifts, then overshoots catastrophically during the merger– and starburst–dominated phase $2\lesssim z\lesssim 5$ (as defined in \citealt{Renaud2021a}), with inferred masses up to a factor $\approx 8$ larger than the dense neutral mass directly measured from the simulation, before settling to a $\approx 50 \%$ underestimate by $z\approx 2$. Even the $L_{\rm [C\,\textsc{ii}]}+Z_{\rm gas}$ prescription is only moderately successful: it stays within $\approx 40 \%$ of the true mass before $z\simeq 5$, but breaks down as soon as the series of major mergers begins, again overpredicting $M_{\rm DN}$ by large factors in the $2\lesssim z\lesssim 5$ window. These failures reflect the strong, merger–driven evolution of the multi-phase ISM during this epoch: the fraction of [C\,\textsc{ii}] arising from non-dense neutral gas changes rapidly, and the same pair $(L_{\rm [C\,\textsc{ii}]},Z_{\rm gas})$ can correspond to very different amounts of dense neutral gas. This confirms the special role of mergers in setting the emission of tracers of the dense ISM, as already noted for CO at low redshift by \cite{Renaud2019}. As a result, no single, redshift-independent global prescription $\alpha_{\rm [C\,\textsc{ii}]}(L_{\rm [C\,\textsc{ii}]},Z_{\rm gas})$ or $\alpha_{\rm [C\,\textsc{ii}]}(Z_{\rm gas})$ can recover the dense neutral gas mass across the full evolutionary history sampled here, suggesting that such prescriptions are unlikely to remain accurate for galaxies whose mass and metallicity evolve strongly with redshift.

Overall, Fig.~\ref{fig:mass_inference_bias} shows that [C\,\textsc{ii}] emission in \textsc{Vintergatan} arises predominantly from regions that also contain dense neutral gas. In addition, resolved prescriptions based on $\Sigma_{\rm [C\,\textsc{ii}]}$ and $Z_{\rm gas}$ recover the total dense neutral mass to within factors of order unity, while global prescriptions that depend on $L_{\rm [C\,\textsc{ii}]}$ and/or the integrated metallicity fail, especially during the merger-dominated epoch. What these fits and the redshift evolution of Fig.~\ref{fig:alpha_evolution} do not explain is why $\alpha_{\rm [C\,\textsc{ii}]}$ varies so strongly with redshift and dynamical state, or which interstellar phases are responsible for the excess or deficit of [C\,\textsc{ii}] emission at a given dense neutral gas mass. To address these questions, we now turn to the 3D phase structure of the ISM and its projections. We examine how [C\,\textsc{ii}] emission populates density–temperature phase space and how mergers and rapid enrichment episodes redistribute [C\,\textsc{ii}] between dense and diffuse neutral, and ionised gas phases.

\begin{figure*}[ht]
\centering
\includegraphics{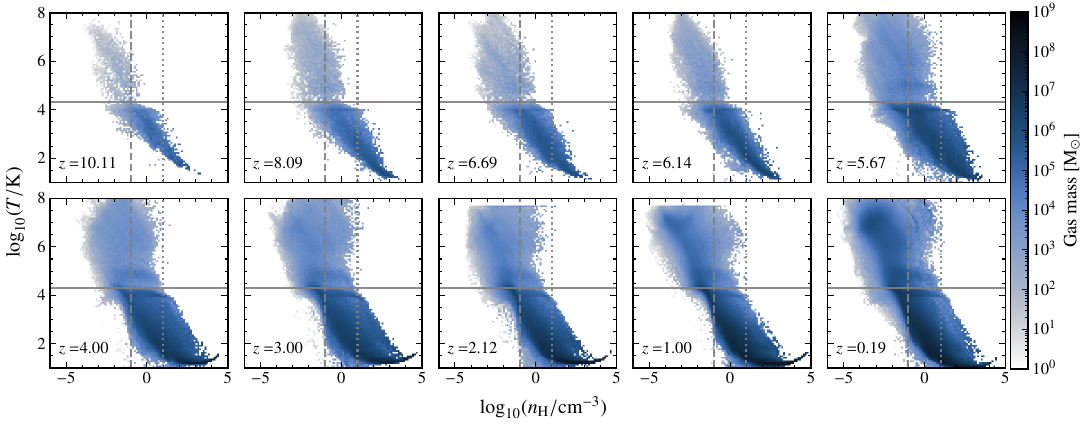}
\caption{Phase diagram of the total gas contents in \textsc{Vintergatan}. The horizontal solid line marks the temperature of $T=2\times10^4~{\rm K}$ used to define cold gas. Dashed and dotted vertical lines represent the density of $0.1~\mathrm{cm^{-3}}$ and $10~\mathrm{cm^{-3}}$ adopted to identify neutral and dense neutral gas, respectively.}
\label{fig:3DPhase}
\end{figure*}

\subsection{Physical origin of [CII] across cosmic time}
\label{subsec:physics_transition}

To interpret the behaviour of the [C\,\textsc{ii}]–to–dense neutral gas conversion factors, we now examine how the phase structure of the ISM evolves over cosmic time, and then relate this to the origin of the [C\,\textsc{ii}] emission. At the earliest times ($z\gtrsim 8$), both the diffuse and dense neutral reservoirs remain modest according to our phase definitions (Fig.~\ref{fig:3DPhase}). This reflects the combined effects of very low metallicity and inefficient or absent dust shielding, which prevent the gas from cooling and settling into the high-density, low-temperature regime required for abundant H$_2$ formation. As the galaxy evolves towards $z\approx 6$, both the diffuse and dense neutral gas masses grow substantially. At $z=7.5$, the diffuse neutral phase becomes a significant, and for $\approx 250$ Myr even slightly dominant, fraction of the cold gas reservoir, while the dense neutral component increases but still occupies a relatively small volume and fraction of the total mass. The phase diagram shows that a large fraction of the gas has cooled below $2\times10^4~\rm K$ but remains below the density threshold of the dense neutral phase.

\begin{figure*}
\centering
\includegraphics{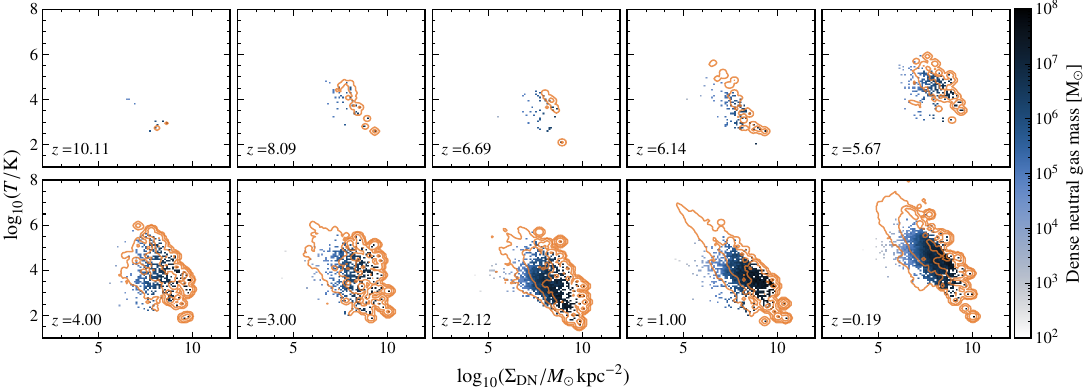}
\caption{Phase diagrams of the 2D-projected dense neutral gas. The background shows the distribution of gas that satisfies our dense neutral-phase criteria in 3D ($T<2\times10^{4}$ K, $n>10$ cm$^{-3}$), projected along the line of sight and binned in effective surface density and temperature. Orange contours show the [C\,\textsc{ii}] emission across all gas phases.}
\label{fig:2Dphase_diagrams}
\end{figure*}

The connection to [C\,\textsc{ii}] is illustrated by the projected temperature--surface-density diagrams of the dense neutral gas shown in Fig.~\ref{fig:2Dphase_diagrams}, where we overplot the [C\,\textsc{ii}]-emitting regions as contours. Because these diagrams are based on projected quantities, a given pixel can mix gas at different 3D temperatures and densities, and is assigned an effective temperature that can lie above the $T<2\times10^{4}$ K threshold used to define the dense neutral phase. This does not imply that genuinely hot gas is classified as dense neutral; the phase definition is applied in 3D before projection. At the earliest times, [C\,\textsc{ii}] emission is very weak and the contours occupy only the coldest, densest tip of the dense neutral-gas distribution, corresponding to the compact regions where the metallicity is highest. The total dense neutral-gas mass is nevertheless small and strongly concentrated in these peaks, such that [C\,\textsc{ii}] already overlaps with most of the dense neutral phase while leaving most of the diffuse neutral reservoir untraced. As enrichment proceeds and the neutral reservoir grows towards $z\approx 6$, the [C\,\textsc{ii}] emission becomes progressively stronger and spreads over a larger fraction of the cold neutral gas distribution. At these redshifts, a non-negligible part of the [C\,\textsc{ii}] luminosity arises from gas that does not meet our dense neutral-gas criterion, while parts of the dense neutral phase remain outside the [C\,\textsc{ii}] contours. Thus, [C\,\textsc{ii}] traces both dense and diffuse neutral gas, but does not yet coincide with the full dense neutral-gas reservoir.

\begin{figure}
\centering
\includegraphics{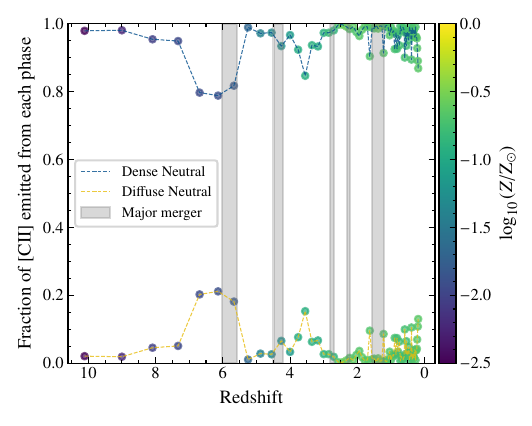}
\caption{Fractional [C\,\textsc{ii}] contribution from the neutral gas phases as a function of redshift. For each snapshot we compute the [C\,\textsc{ii}] luminosity emitted by the dense and diffuse neutral phases separately and plot, as filled circles, the fraction of the total [C\,\textsc{ii}] luminosity that arises from each phase. Points are colour-coded by the global gas metallicity. Vertical grey bands indicate the redshift extent of major 1:1 merger events.}
\label{fig:tracer_fraction}
\end{figure}

This evolution in the relative contributions of the diffuse and dense neutral phases fundamentally changes what the [C\,\textsc{ii}] line traces. Figure~\ref{fig:tracer_fraction} shows the fractional contribution of these phases to the total [C\,\textsc{ii}] luminosity as a function of redshift. At very high redshift ($z \gtrsim 8$), the dense neutral reservoir is still small and much of the cold gas is concentrated in the densest regions; even at low metallicity, the [C\,\textsc{ii}] emission then arises primarily from the high-density peaks of the cold ISM. During the subsequent build-up of dense neutral gas ($6 \lesssim z \lesssim 8$), as metallicity increases and the neutral reservoir grows, [C\,\textsc{ii}] emission traces an increasing fraction of the neutral gas. Around $z \approx 5.7$-6, coincident with the first major merger experienced by \textsc{Vintergatan}, the metallicity reaches a level at which dust shielding becomes efficient over a substantial fraction of the disc. The dense neutral phase then becomes the dominant cold component, while [C\,\textsc{ii}] continues to arise from a wide range of densities (Fig.~\ref{fig:tracer_fraction}). In this regime, the [C\,\textsc{ii}]-bright region in phase space closely follows the locus of the cold, dense neutral gas, and essentially all of the dense neutral mass lies within this region. [C\,\textsc{ii}] therefore becomes a nearly complete tracer of the dense neutral phase.

From $z \approx 5$ down to $z \approx 0$, the dense neutral phase remains the dominant origin of the [C\,\textsc{ii}] emission, with its contribution staying at $\gtrsim 80\%$ (Fig.~\ref{fig:tracer_fraction}). Over this redshift range, the multiphase structure of the ISM and the local conditions under which [C\,\textsc{ii}] is emitted keep evolving, but the dominant [C\,\textsc{ii}]-emitting phase does not change. An increasing fraction of the dense neutral gas resides in environments where [C\,\textsc{ii}] is relatively faint, and a small part of this phase becomes effectively [C\,\textsc{ii}]-dark (blue pixels outside the orange contours in the second row of Fig.~\ref{fig:2Dphase_diagrams}). The fractional [C\,\textsc{ii}] contributions from dense and diffuse neutral gas vary only weakly with time (Fig.~\ref{fig:tracer_fraction}), indicating that mergers and starburst episodes primarily modulate the excitation conditions and $\alpha_{\rm [C\,\textsc{ii}]}$, rather than modifying the dominant [C\,\textsc{ii}]-emitting phase. The earliest major merger at $z\simeq 6$ coincides with one of the largest excursions in the fractional [C\,\textsc{ii}] contributions from the dense and diffuse neutral phases (Fig.~\ref{fig:tracer_fraction}). Around this epoch, the galaxy is still rapidly assembling its cold gas reservoir and undergoing strong chemical enrichment; the merger compresses the gas and accelerates the build-up of dense, enriched regions, which amplifies the [C\,\textsc{ii}] output of the dense neutral phase. We also note that a rotationally supported disc emerges around $z\simeq 5$ \citep[their Fig.~5]{SegoviaOtero2022}. Although one might expect a sharp change in dust attenuation or UV shielding at disc settling to leave a clear imprint on $\alpha_{\rm [C\,\textsc{ii}]}$, we do not identify any such direct causal link. Instead, the temporal coincidence suggests that the evolution of $\alpha_{\rm [C\,\textsc{ii}]}$ is more likely an indirect consequence of disc settling redistributing carbon in phase space, rather than a sudden change in the underlying [C\,\textsc{ii}] physics. Later major mergers at $z\lesssim 6$ occur once the ISM is already metal-rich and dominated by dense neutral gas, and induce only modest ($\lesssim 20\%$) fluctuations in the phase fractions. We therefore interpret the pronounced feature at $z\approx 6$ in Fig.~\ref{fig:tracer_fraction} as the combined result of the merger and the underlying secular evolution of gas surface density and metallicity, rather than as an isolated causal imprint of the merger alone. More generally, the excursions in the phase fractions around merger epochs remain brief (a few hundred Myr), suggesting that mergers drive short-lived perturbations in the [C\,\textsc{ii}] emission from the different neutral gas phases.

In summary, the evolution of the phase diagram provides a physical explanation for the behaviour of the $\alpha_{\rm [C\,\textsc{ii}]}$ prescriptions discussed in Sect.~\ref{subsec:prescriptions_relations}. At very high redshift, large $\alpha_{\rm [C\,\textsc{ii}]}$ values arise because [C\,\textsc{ii}] emission is confined to dense, metal-poor regions, rather than because the dense neutral reservoir is spatially untraced by [C\,\textsc{ii}]. During the build-up phase, [C\,\textsc{ii}] increasingly traces the full neutral envelope rather than only the dense neutral component. Global calibrations that assume a fixed phase mix can therefore associate [C\,\textsc{ii}] luminosity with substantially different amounts of dense neutral gas at different epochs. Once the ISM is enriched and [C\,\textsc{ii}] emission is dominated by the dense neutral phase (Fig.~\ref{fig:tracer_fraction}), mergers and rapid chemical evolution between $2\lesssim z\lesssim 5$ drive strong evolution in the density, metallicity, and radiation field of the [C\,\textsc{ii}]-emitting gas. This is precisely the redshift range where single global prescriptions $\alpha^{\rm glob}_{\rm [C\,\textsc{ii}]}(L_{\rm [C\,\textsc{ii}]},Z_{\rm gas})$ or $\alpha^{\rm glob}_{\rm [C\,\textsc{ii}]}(Z_{\rm gas})$ produce order-of-magnitude biases in the inferred dense neutral-gas mass. Because the resolved prescriptions are calibrated locally, they better capture the pixel-to-pixel variations in the physical conditions of the [C\,\textsc{ii}]-emitting dense neutral gas than a single global $\alpha_{\rm [C\,\textsc{ii}]}$ tied only to integrated $L_{\rm [C\,\textsc{ii}]}$ and $Z_{\rm gas}$.

\section{Discussion}\label{sec:discussion}
\subsection{Caveats and limitations}

Our analysis is based on a single high-resolution cosmological zoom-in simulation of a Milky Way-like progenitor, and therefore follows one specific mass assembly and merger history rather than a statistical population of galaxies. While the ISM conditions in \textsc{Vintergatan} broadly resemble those inferred for main-sequence systems of comparable mass at $z \lesssim 3$ \citep[e.g.][]{Tacconi2018}, the galaxy is substantially less massive and more metal-poor than the observed luminous, [C\,\textsc{ii}]-bright systems that dominate current high-redshift samples \citep[e.g.][]{HerreraCamus2025}. This work should therefore be viewed not as a universal calibration but as a detailed example of how [C\,\textsc{ii}] traces the ISM in a galaxy that evolves from a low-mass, metal-poor progenitor into a present-day disc. The mechanisms we identify, which link $\alpha_{\rm [C\,\textsc{ii}]}$ to phase mix, metallicity, and assembly history, are expected to operate in a broad range of systems; the resulting trends and caveats should therefore remain relevant well beyond this particular Milky Way analogue.

A second limitation arises from the adopted sub-grid and radiative transfer prescriptions. The \textsc{Skirt}/\textsc{Toddlers} implementation relies on pre-computed photoionisation and PDR models, and assumes a fixed dust-to-metal ratio as a function of metallicity. At very low metallicity, both the dust content and the covering fraction of dense PDRs are poorly constrained, which could introduce systematic uncertainties in the predicted [C\,\textsc{ii}] emissivity and phase mix, especially at $z \gtrsim 8$. Likewise, our definition of the dense neutral phase is based on gas density and temperature cuts rather than an explicit chemical network; it should therefore be regarded as a proxy for, rather than a direct measurement of, the molecular reservoir. Physically motivated prescriptions for the H$_2$ fraction that account for shielding and metallicity \citep[e.g.][]{Krumholz2009} and detailed non-equilibrium networks for CO and related species \citep[e.g.][]{Glover2010_CO} provide more physical treatments of the atomic-to-molecular transition and CO chemistry than our simple density- and temperature-based phase definition. However, they involve large reaction networks with many density- and dust-dependent channels and typically require unresolved clumping factors and other sub-grid assumptions that remain uncertain at the resolution of galaxy-scale simulations. Given these uncertainties, and our focus on [C\,\textsc{ii}] rather than CO, we adopted this simple density- and temperature-based definition as a pragmatic proxy for the dense gas expected to host the H$_2$ reservoir. Our conclusions therefore concern the association between [C\,\textsc{ii}] emission and dense neutral gas; the exact correspondence of this phase with the molecular reservoir depends on the adopted phase definition.

More sophisticated treatments that evolve dust and follow non-equilibrium chemistry for H$_2$, CO, C and [C\,\textsc{ii}] have recently been used to study high-redshift galaxies. For instance, the \textsc{ColdSIM} simulations include time-dependent chemistry and show that [C\,\textsc{ii}] emission is dominated by diffuse neutral gas but remains a robust tracer of the molecular gas mass, with conversion factors that vary systematically with metallicity and star formation rate but do not exhibit order-of-magnitude swings at fixed $M_{\rm mol}$ \citep{Casavecchia2025}. Similarly, the MARIGOLD simulations, which use the HYACINTH non-equilibrium chemistry network, find that accounting for secondary dependencies in the $L_{\rm [C\,\textsc{ii}]}$-$M_{\rm mol}$ relation improves $M_{\rm mol}$ predictions by a factor of $\sim 2.3$ \citep{Khatri2025}. These studies provide a quantitative benchmark for the uncertainty introduced when relating [C\,\textsc{ii}] to molecular gas. While they resolve the multiphase ISM on $\lesssim 10$-50 pc scales with on-the-fly chemistry, compared to our $\sim 100$ pc post-processed grid, they suggest that dust and chemistry can modify [C\,\textsc{ii}]-to-molecular-gas conversion factors at the factor-of-few level. This comparison should nevertheless be interpreted with caution, since our conversion factor is defined relative to the dense neutral phase rather than an explicitly modelled H$_2$ mass.

Despite these caveats, several of our conclusions are qualitatively robust against plausible variations in the sub-grid physics and radiative transfer details. The close correspondence between global and resolved $\alpha_{\rm [C\,\textsc{ii}]}$, the near-unity recovered dense neutral-gas mass fractions after the dense neutral phase becomes dominant, and the strong redshift evolution of global $\alpha_{\rm [C\,\textsc{ii}]}$ all persist when varying the spatial aperture used to extract the emission, projection, and possible [C\,\textsc{ii}] self-absorption within reasonable bounds (Appendix~\ref{apdx:left_out}). These features reflect generic changes in ISM structure, metallicity, and surface density along the \textsc{Vintergatan} track, rather than being driven by a small subset of extreme regions or by a specific choice of modelling parameters. Extending this analysis to suites of zoom-in simulations spanning a wider range of stellar masses and enrichment histories, and explicitly following molecular chemistry, is essential to quantify how the characteristic transition redshift, the [C\,\textsc{ii}]-to-dense-neutral-gas relation, and its connection to the molecular reservoir, scale with mass and metallicity.

\subsection{Implications for [CII] as a molecular gas tracer}

From an observational perspective, our results emphasise both the strengths and limitations of [C\,\textsc{ii}] as a tracer of the dense neutral ISM and, by extension, of the molecular reservoir at high redshift. In \textsc{Vintergatan}, [C\,\textsc{ii}] is almost always co-spatial with the dense neutral phase once the galaxy has reached metallicities typical of $z \approx 2$-5 main-sequence systems, and resolved prescriptions based on $\Sigma_{\rm [C\,\textsc{ii}]}$ and $Z_{\rm gas}$ recover its total mass to within factors of order unity. Since this phase encompasses the gas expected to host the H$_2$ reservoir, these results support the use of [C\,\textsc{ii}] as a practical molecular-gas tracer in enriched systems, provided that the correspondence between dense neutral and molecular gas is calibrated appropriately. At earlier times ($6 \lesssim z \lesssim 8$), during the merger-dominated assembly phase, $\alpha_{\rm [C\,\textsc{ii}]}$ evolves rapidly and [C\,\textsc{ii}] increasingly traces the total neutral gas reservoir rather than only its densest component. Our results therefore caution against applying low-redshift, globally calibrated $\alpha_{\rm [C\,\textsc{ii}]}$ relations to low-mass systems at high redshift without accounting for their lower metallicities and rapidly evolving conditions.

For current ALMA observations of massive $z \approx 4\text{--}7$ galaxies, which typically probe systems more massive and more [C\,\textsc{ii}]-luminous than \textsc{Vintergatan} at the same redshift, our findings support [C\,\textsc{ii}] as a useful tracer of the dense gas associated with the molecular reservoir. However, the adopted $\alpha_{\rm [C\,\textsc{ii}]}$ or scaling relation should match the regime in which it was calibrated, in particular in stellar mass and metallicity. 
In this context, spatially resolved analyses that combine $\Sigma_{\rm [C\,\textsc{ii}]}$ with metallicity information are more reliable than a single global conversion factor, because they better capture variations in [C\,\textsc{ii}] emissivity across the evolving ISM. 
Ongoing surveys such as ALMA–CRISTAL, which map [C\,\textsc{ii}] and star formation on kiloparsec scales in main-sequence galaxies at $4 \leq z \leq 6$ \citep[e.g.][]{Li2024,Accard2025}, already reveal [C\,\textsc{ii}] emission that is more extended than the rest-frame UV and FIR continuum and enable spatially resolved [C\,\textsc{ii}]\text{--}SFR and Kennicutt–Schmidt relations at $z \gtrsim 4$. Our results provide a framework for interpreting such data and argue that adopting resolved, $\Sigma_{\rm [C\,\textsc{ii}]}$- and metallicity-dependent $\alpha_{\rm [C\,\textsc{ii}]}$ prescriptions, rather than a single global value, is crucial for robust estimates of the dense neutral gas associated with the molecular reservoir.

Looking ahead to less massive systems and to future facilities, our results motivate tests of how [C\,\textsc{ii}]-based gas estimates depend on redshift, stellar mass, metallicity, and merger events, rather than relying on a single, universally applicable $\alpha_{\rm [C\,\textsc{ii}]}$. In particular, the strong rise of $\alpha_{\rm [C\,\textsc{ii}]}$ in the very metal-poor regime suggests that lower-mass galaxies, which typically enrich more slowly than Milky Way progenitors, may remain [C\,\textsc{ii}]-deficient down to later times and thus require systematically higher conversion factors at fixed $L_{\rm [C\,\textsc{ii}]}$ than the massive, [C\,\textsc{ii}]-bright systems targeted so far. Systematic comparisons between high-resolution ($\lesssim 50\,{\rm pc}$) simulations of such low-mass galaxies and forthcoming deep [C\,\textsc{ii}] surveys are crucial to quantify the evolutionary patterns identified here in the case of other systems.

\section{Conclusions}\label{sec:conclusion}

We combine the high-resolution cosmological zoom-in simulation \textsc{Vintergatan} with the \textsc{Skirt} radiative transfer code to follow the joint evolution of [C\,\textsc{ii}] emission and the dense neutral ISM in a Milky Way--like progenitor from $z\simeq10$ to $z\simeq0.2$. Our dense neutral phase, defined through temperature and density thresholds, is used as a proxy for the molecular reservoir but does not constitute an explicit H$_2$ measurement.

The galaxy-integrated $L_{\rm [C\,\textsc{ii}]}$--$M_{\rm DN}$ and $L_{\rm [C\,\textsc{ii}]}$--SFR relations evolve strongly with redshift, primarily driven by chemical enrichment and changes in the distribution of dense neutral gas within the galaxy. At very early times, when the galaxy is still low-mass and metal-poor, the system follows a steep, [C\,\textsc{ii}]-deficient sequence: $L_{\rm [C\,\textsc{ii}]}$ and $L_{\rm [C\,\textsc{ii}]}/L_{\rm IR}$ lie well below local and high-redshift calibrations, and the conversion factor $\alpha_{\rm [C\,\textsc{ii}]}$ reaches $\gtrsim 10^3\,{\rm M_\odot\,L_\odot^{-1}}$. Once the ISM reaches $Z_{\rm gas} \gtrsim 0.05$--$0.1\,{\rm Z_\odot}$ at $z\lesssim 5$, these relations become nearly linear and approach empirically calibrated relations for chemically evolved galaxies, with $\alpha_{\rm [C\,\textsc{ii}]} \sim 10$--$10^2\,{\rm M_\odot\,L_\odot^{-1}}$.

The large range in $\alpha_{\rm [C\,\textsc{ii}]}$ does not arise because most of the dense neutral gas becomes [C\,\textsc{ii}]-dark; instead, it reflects how efficiently gas in this phase radiates in [C\,\textsc{ii}] under different metallicities, phase mixes, and merger events. Throughout the evolution, dense neutral gas contained in [C\,\textsc{ii}]-emitting pixels accounts for $\sim90$--$100$ per cent of the total $M_{\rm DN}$ at all redshifts, with larger deviations only at the latest times when a small fraction of the dense neutral gas becomes effectively [C\,\textsc{ii}]-dark. In the extremely metal-poor regime, low dense neutral-gas masses combined with metal-enriched outflows reduce the carbon abundance in the cold ISM, leading to faint [C\,\textsc{ii}] emission per unit dense neutral-gas mass even when $M_{\rm DN}$ is non-negligible.

Both global and spatially resolved estimates of $\alpha_{\rm [C\,\textsc{ii}]}$ carry different information. The global conversion factor, obtained from the ratio of total $M_{\rm DN}$ to total $L_{\rm [C\,\textsc{ii}]}$, tracks the median of the resolved pixel distribution at $z\gtrsim 6$, indicating a coherent shift of the entire distribution of local $\alpha_{\rm [C\,\textsc{ii}]}$ values with redshift. At later times, however, the global $\alpha_{\rm [C\,\textsc{ii}]}$ becomes increasingly dominated by a small number of bright [C\,\textsc{ii}] regions with moderate local conversion factors, while numerous fainter pixels populate a high-$\alpha_{\rm [C\,\textsc{ii}]}$ tail. The difference between the global value and the median of the resolved values therefore encodes how [C\,\textsc{ii}] and the dense neutral gas are redistributed within the galaxy as it forms a metal-rich disc. Across the full history of \textsc{Vintergatan}, the effective $\alpha_{\rm [C\,\textsc{ii}]}$ spans nearly three orders of magnitude, and no single, redshift-independent global value can represent the underlying local conditions.

We provide practical prescriptions relating [C\,\textsc{ii}] emission to the dense neutral phase. Resolved relations of the form $\alpha_{\rm [C\,\textsc{ii}]}(\Sigma_{\rm [C\,\textsc{ii}]},Z_{\rm gas})$, calibrated on 100 pc pixels, recover the total dense neutral-gas mass within factors of order unity over most of the redshift range, with deviations that correlate primarily with major mergers and rapid enrichment episodes. In contrast, global prescriptions depending only on $L_{\rm [C\,\textsc{ii}]}$ and/or integrated metallicity can under- or over-estimate $M_{\rm DN}$ by up to an order of magnitude during merger-dominated phases. These results imply that spatially resolved analyses, which exploit $\Sigma_{\rm [C\,\textsc{ii}]}$ together with metallicity information, are substantially more reliable than unique global conversion factors for tracing the dense neutral ISM.

Examining the phase diagram provides a coherent physical explanation for the behaviour of $\alpha_{\rm [C\,\textsc{ii}]}$. At very high redshift, [C\,\textsc{ii}] emission is limited to dense, metal-poor peaks, leaving most of the diffuse neutral reservoir untraced. During the build-up of the dense neutral phase, [C\,\textsc{ii}] progressively extends over the neutral gas envelope and draws an increasing contribution from diffuse neutral gas as the dense-gas fraction and metallicity grow. Around $z\simeq 5.7$--6, in coincidence with the first major merger and the onset of efficient dust shielding over a substantial fraction of the disc, [C\,\textsc{ii}] becomes an almost complete tracer of the dense neutral phase, and the global $\alpha_{\rm [C\,\textsc{ii}]}$ remains close to the median of the resolved distribution. At later times ($z\lesssim 4$--3), the ISM becomes more structured: the global $\alpha_{\rm [C\,\textsc{ii}]}$ is increasingly set by a small number of bright [C\,\textsc{ii}] regions with moderate conversion factors, while numerous fainter pixels populate a high-$\alpha_{\rm [C\,\textsc{ii}]}$ tail. The difference between the global value and the median of the resolved values thus encodes how [C\,\textsc{ii}] emission and dense neutral gas are redistributed within the galaxy as it becomes more metal-rich. This evolution explains why global prescriptions $\alpha_{\rm [C\,\textsc{ii}]}(L_{\rm [C\,\textsc{ii}]},Z)$ or $\alpha_{\rm [C\,\textsc{ii}]}(Z_{\rm gas})$ calibrated at later times can fail when extrapolated into the low-metallicity, merger-dominated regime.

Finally, our analysis identifies the conditions under which [C\,\textsc{ii}] can reliably trace the dense neutral ISM and can therefore serve as a useful proxy for the molecular reservoir. Once metallicities and masses comparable to $z\approx2$--5 main-sequence galaxies are reached, and particularly in systems more enriched and more [C\,\textsc{ii}]-luminous than \textsc{Vintergatan}, appropriately calibrated, regime-dependent $\alpha_{\rm [C\,\textsc{ii}]}$ prescriptions should provide useful constraints on the gas associated with the molecular reservoir. At earlier times and in lower-mass, less metal-rich galaxies, the strong dependence of $\alpha_{\rm [C\,\textsc{ii}]}$ on metallicity, ISM phase mix, and mergers implies that blindly applying a unique conversion factor derived from low-redshift data can introduce order-of-magnitude biases. Our results provide a physically motivated template for Milky Way--mass progenitors and a framework for constructing mass- and metallicity-dependent [C\,\textsc{ii}]-to-dense-neutral-gas conversion functions that can be confronted with deeper [C\,\textsc{ii}] observations jointly analysed with JWST rest-frame optical spectroscopy, which now delivers robust gas-phase metallicities and spatially resolved abundance gradients in high redshift galaxies.

\begin{acknowledgements}
CA acknowledges that this work of the Interdisciplinary Thematic Institute IRMIA++, as part of the ITI 2021-2028 programme of the University of Strasbourg, CNRS and Inserm, was supported by IdEx Unistra (ANR-10-IDEX-0002), and by SFRI-STRAT’US project (ANR-20-SFRI-0012) under the framework of the French Investments for the Future Programme.
OA acknowledges support from the Knut and Alice Wallenberg Foundation, the Swedish Research Council (grant 2025-04892), the Swedish National Space Agency (SNSA Dnr 2023-00164), the LMK Foundation, and eSSENCE.
We acknowledge the use of Astropy \citep{Astropy}, Matplotlib \citep{Matplotlib}, Scipy \citep{Scipy}.\end{acknowledgements}

\bibliographystyle{aa}
\bibliography{biblio}

\begin{appendix}

\section{Orientation, extraction radius, timescale, and transparency effects}
\label{apdx:left_out}

In this appendix we quantify how viewing angle, extraction aperture, and line opacity affect the inferred [C\,\textsc{ii}]–to–gas relations. We remind the reader that unless stated otherwise, all measurements in the main text use the face-on view, an aperture of three stellar half-mass radii, $3\,r_{\rm H}$, and the fully radiative-transferred [C\,\textsc{ii}] luminosities.

\begin{figure}
\centering
\includegraphics[width=0.992\linewidth]{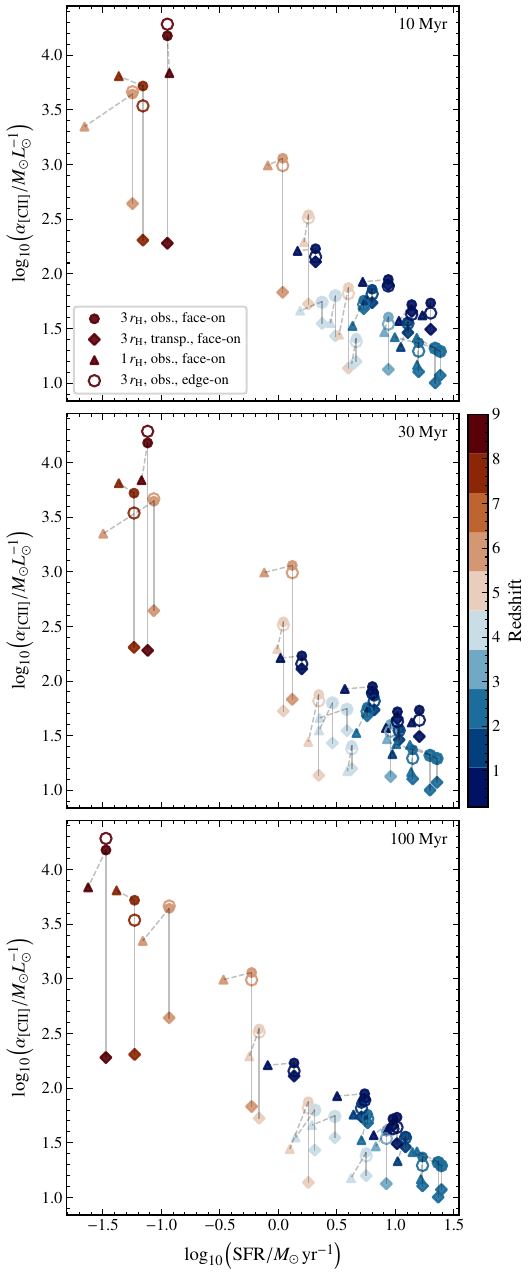}
\caption{Global $\alpha_{\rm [CII]}$ as a function of SFR for three star-formation averaging timescales, colour-coded by redshift. Filled circles show the fiducial measurements (face-on, $3\,r_{\rm H}$, observed [C\,\textsc{ii}] luminosities), open circles show the corresponding edge-on measurements, diamonds indicate the values obtained from the intrinsic transparent [C\,\textsc{ii}] luminosities, and triangles show the observed measurements within $1\,r_{\rm H}$. Grey solid segments connect the observed and transparent measurements at fixed $3\,r_{\rm H}$, while grey dashed segments connect the $3\,r_{\rm H}$ and $1\,r_{\rm H}$ measurements at fixed redshift.}
\label{fig:alphaCII_SFR_appendix}
\end{figure}

Figure~\ref{fig:alphaCII_SFR_appendix} shows the global $\alpha_{\rm [CII]}$–SFR relation for three star-formation averaging timescales, colour-coded by redshift. Filled circles correspond to our fiducial configuration (face-on, $3\,r_{\rm H}$, observed [C\,\textsc{ii}] luminosities). For each snapshot, diamonds at the same SFR indicate the values obtained from the intrinsic transparent [C\,\textsc{ii}] luminosities within $3\,r_{\rm H}$, and triangles show the observed measurements within a smaller aperture of $1\,r_{\rm H}$. Open circles mark the $3\,r_{\rm H}$, edge-on measurements, which can be compared directly to their filled counterparts at the same redshift. Grey solid segments connect the observed and transparent measurements at fixed $3\,r_{\rm H}$, while grey dashed segments connect the $3\,r_{\rm H}$ and $1\,r_{\rm H}$ measurements at fixed redshift. The overall locus becomes marginally tighter as the SFR averaging timescale increases from 10 to 100 Myr, but this effect remains small compared to the intrinsic scatter.

The dashed segments demonstrate that changing the extraction radius from $3\,r_{\rm H}$ to $1\,r_{\rm H}$ has only a modest impact on the inferred conversion factor. At fixed SFR and redshift, the $1\,r_{\rm H}$ points lie systematically below the $3\,r_{\rm H}$ fiducial ones by at most $\approx 0.1$–$0.2$\,dex, reflecting the higher fraction of dense neutral, metal-rich gas enclosed by the smaller aperture. The open and filled circles nearly overlap at all redshifts, showing that face-on and edge-on views yield very similar $\alpha_{\rm [CII]}$ values, with orientation effects well below the intrinsic scatter of the relation.

The solid segments highlight the impact of [C\,\textsc{ii}] self-absorption. At $z \lesssim 5$, transparent and fully radiative-transferred measurements agree to within a few tens of per cent, indicating that line opacity is negligible once the galaxy is more extended and the ISM is less compact. At earlier times ($z \gtrsim 6$), when the system is in a dense assembly phase, the transparent points move to somewhat lower $\alpha_{\rm [CII]}$ (higher $L_{\rm [CII]}$) than their radiative-transferred counterparts, consistent with self-absorption suppressing the emergent [C\,\textsc{ii}] flux by up to a factor of a few. Even in this regime, however, the qualitative form of the [C\,\textsc{ii}]–dense neutral gas relation and the strong redshift evolution of $\alpha_{\rm [CII]}$ remain unchanged. We therefore conclude that our main results are robust against reasonable variations in orientation, aperture choice, SFR averaging timescale, and [C\,\textsc{ii}] line opacity.

\newpage
\section{Global and resolved \texorpdfstring{$\alpha_{[\mathrm{CII}]}$}{alpha\_[CII]} as a function of metallicity}
\label{app:alpha_vs_Z}

\begin{figure}[h]
\centering
\includegraphics[width=0.9\linewidth]{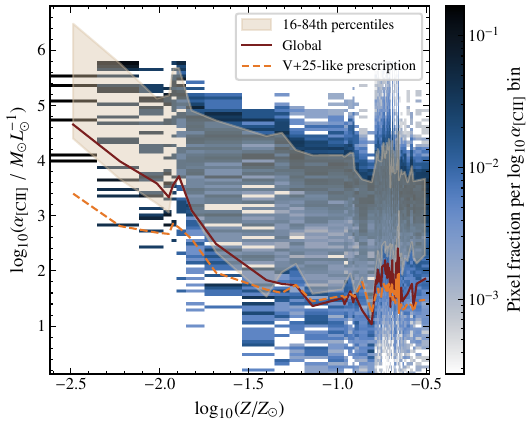}
\caption{Same as Fig.~\ref{fig:alpha_evolution}, but as a function of the gas metallicity. The blue colour map shows the distribution of resolved $\alpha_{\rm [C\,\textsc{ii}]}$ values, the dark red line the global conversion factor, and the shaded region the 16th–84th percentile range of the resolved values. The dashed line marks the metallicity-dependent $\alpha_{\rm [C\,\textsc{ii}]}$ range from \citet{Vallini2025}.}
\label{fig:alpha_vs_Z}
\end{figure}

In Fig.~\ref{fig:alpha_vs_Z} we show the same information as in Fig.~\ref{fig:alpha_evolution}, but now as a function of the global, gas-mass-weighted metallicity instead of redshift. Each vertical slice corresponds to one simulation snapshot and is placed at its \(\log_{10}(Z_{\rm gas}/{\rm Z_\odot})\). The colour map traces the distribution of resolved \(\alpha_{\rm [C\,\textsc{ii}]}\) values measured on $100~\mathrm{pc}$ pixels, while the shaded region and solid line indicate, respectively, the 16th–84th percentile range and the global conversion factor.

The behaviour of \(\alpha_{\rm [C\,\textsc{ii}]}\) in this plane confirms that the strong evolution seen in Fig.~\ref{fig:alpha_evolution} is primarily controlled by the build-up of metals. At very low metallicities, \(\log_{10}(Z_{\rm gas}/{\rm Z_\odot}) \lesssim -2\), both the global and resolved \(\alpha_{\rm [C\,\textsc{ii}]}\) values are large (\(\gtrsim 10^3~{\rm M_\odot\,L_\odot^{-1}}\)), reflecting the intrinsically faint [C\,\textsc{ii}] emission in the metal-poor ISM. As the ISM enriches and reaches \(\log_{10}(Z_{\rm gas}/{\rm Z_\odot}) \approx -1.3\), the global and median resolved \(\alpha_{\rm [C\,\textsc{ii}]}\) converge towards the nearly constant regime \(\sim 10\)–\(10^2~{\rm M_\odot\,L_\odot^{-1}}\) discussed in Sect.~\ref{subsec:resolved_alpha}. This demonstrates that the transition from the [C\,\textsc{ii}]-deficient to the more standard regime is largely metallicity-driven, and that expressing \(\alpha_{\rm [C\,\textsc{ii}]}\) as a function of \(Z_{\rm gas}\) yields the same qualitative conclusions as the redshift-based view. Additionally, the metallicity-dependent $\alpha_{\rm [C\,\textsc{ii}]} $ range inferred from the \textsc{SERRA} simulations \citep{Vallini2025} broadly overlaps with our enriched regime, but underestimates the steep rise of $\alpha_{\rm [C\,\textsc{ii}]}$ towards very low metallicities.

\newpage

\section{dense neutral-gas mass bias in redshift windows}\label{apdx:bias}

\begin{figure}[h]
  \centering
  \begin{minipage}{0.48\textwidth}
    \centering
    \includegraphics[width=0.895\linewidth]{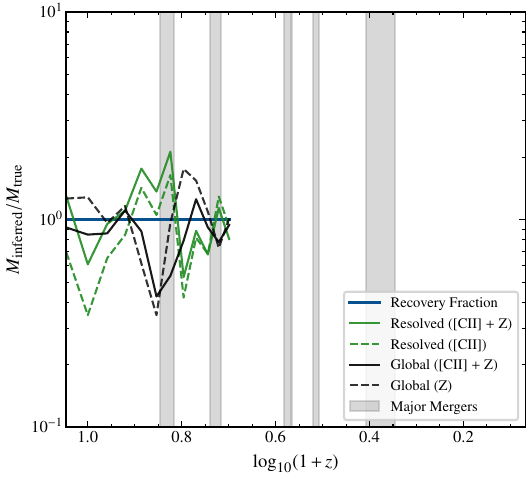}
    \caption{Similar to Fig.~\ref{fig:mass_inference_bias}, fitting only the $z\geq4.00$ snapshots.}
    \label{fig:left}
  \end{minipage}
  \hfill
  \begin{minipage}{0.48\textwidth}
    \centering
    \includegraphics[width=0.895\linewidth]{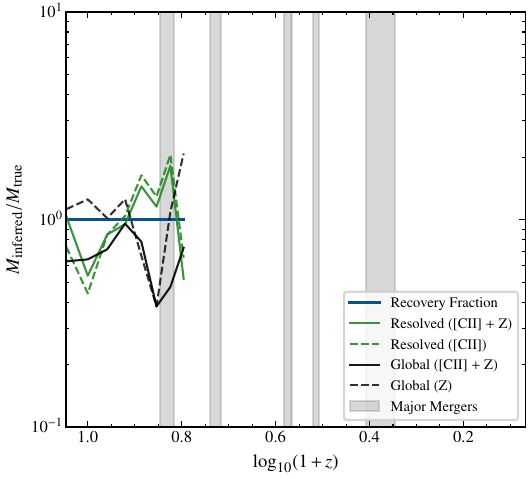}
    \caption{Similar to Fig.~\ref{fig:mass_inference_bias}, fitting only the $z\geq5.25$ snapshots.}
    \label{fig:right}
  \end{minipage}
\end{figure}

In Sect.~\ref{subsec:prescriptions_relations}, we fitted relations for $\alpha_{\rm [C\,\textsc{ii}]}$ using all available snapshots across the full redshift range. To test whether focusing on specific evolutionary stages improves the calibration, we repeated the fits using only snapshots with $z\geq4.0$ and $z\geq5.25$, respectively. Figure~\ref{fig:left} shows the resulting bias in the inferred dense neutral-gas mass when the $z\geq4.0$ relations are applied, while Fig.~\ref{fig:right} presents the same analysis for the $z\geq5.25$ fits.

In these restricted windows, both the global and resolved prescriptions perform better than in the full-sample case. The recovered total dense neutral-gas mass generally lies between $\approx0.6$ and $10$ times the true value, with most snapshots clustering closer to an unbiased estimate. The largest excursions still coincide with major mergers, but the systematic offsets are reduced and the scatter around $M_{\rm DN,inferred}/M_{\rm DN,true}=1$ is noticeably tighter for all four prescriptions. This shows that calibrating $\alpha_{\rm [C\,\textsc{ii}]}$ in narrower, physically more homogeneous redshift ranges can yield more robust global and resolved [C\,\textsc{ii}]-based estimators of the dense neutral-gas mass, in regimes where the ISM structure is not evolving as rapidly as across the full cosmological history.

\end{appendix}

\end{document}